\documentclass[aps,prb,twocolumn,superscriptaddress,showpacs,floatfix,amsmath,amssymb]{revtex4}
\usepackage{graphicx}
\usepackage{color}
\usepackage{xcolor,soul}
\usepackage[normalem]{ulem}

\DeclareMathAlphabet\mathbfcal{OMS}{cmsy}{b}{n}

\newcommand{\br}[0]{ {\bf r} }

\newcommand{\hT}[0]{ \hat{ T } }
\newcommand{\hW}[0]{ \hat{ W } }

\newcommand{\FPd}[0]{ \hat{\Psi}^\dagger }
\newcommand{\FPdr}[0]{ \FPd(\br) }
\newcommand{\FPdur}[0]{ \FPd_\uparrow(\br) }
\newcommand{\FPddnr}[0]{ \FPd_\downarrow(\br) }

\newcommand{\FP}[0]{ \hat{\Psi} }

\newcommand{\SDFT}[0]{ {\rm {\tiny \tiny SDFT} } }
\newcommand{\SCDFT}[0]{ {\rm {\tiny \tiny SCDFT} } }

\newcommand{\KS}[0]{ {\rm {\tiny \tiny KS}} }
\newcommand{\GKS}[0]{ {\rm {\tiny \tiny GKS}} }
\newcommand{\DFA}[0]{ {\rm {\tiny \tiny DFA}} }

\newcommand{\hO}[0]{ \hat{O} }

\newcommand{\vm}[0]{ {\vec m} }

\newcommand{\bj}[0]{ {\bf j} }
\newcommand{\vbJ}[0]{ {\vec {\bf J}} }
\newcommand{\vbA}[0]{ {\vec {\bf A}} }
\newcommand{\bA}[0]{ {\bf A} }

\definecolor{Mygrey}{gray}{0.80}
\definecolor{lteal}{rgb}{0.10,0.60,0.70}
\definecolor{dkred}{rgb}{0.80,0.10,0.00}

\newcommand{\comment}[1]{}

\newcommand*{\LightComments}{}%

\ifdefined\NoComments

\else
 \ifdefined\LightComments

\else

 \fi
\fi

\begin{document}
\author{Jacques K. Desmarais}
\email{jacqueskontak.desmarais@unito.it}
\affiliation{Dipartimento di Chimica, Universit\`{a} di Torino, via Giuria 5, 10125 Torino, Italy}

\author{Giacomo Ambrogio}
\affiliation{Dipartimento di Chimica, Universit\`{a} di Torino, via Giuria 5, 10125 Torino, Italy}

\author{Giovanni Vignale}
\affiliation{Institute for Functional Intelligent Materials, National University of Singapore, 4 Science Drive 2, Singapore 117544}

\author{Alessandro Erba}
\affiliation{Dipartimento di Chimica, Universit\`{a} di Torino, via Giuria 5, 10125 Torino, Italy}

\author{Stefano Pittalis}
\email{stefano.pittalis@nano.cnr.it}
\affiliation{Istituto Nanoscienze, Consiglio Nazionale delle Ricerche, Via Campi 213A, I-41125 Modena, Italy}

\title{Generalized Kohn-Sham Approach for the Electronic Band Structure of Spin-Orbit Coupled Materials}

\date{\today}

\begin{abstract}
Spin-current density functional theory (SCDFT) is a formally exact framework designed to handle the treatment of interacting many-electron systems including spin-orbit coupling (SOC) at the level of the Pauli equation. In practice,  robust and accurate calculations of the electronic structure of these systems call for functional approximations that depend not only on the densities and currents 
but also on spinors explicitly.
Here we extend the generalized Kohn-Sham (GKS) approach of [{\em Seidl et all. ``Generalized Kohn-Sham schemes and the band-gap problem'', Phys. Rev. B {\bf 53}, 3764 (1996)}] to SCDFT.  
This framework entails the prominent cases of  hybrid forms and meta-generalized-gradient-approximations.
We clarify that the  exchange-correlation potentials conjugate to the currents 
need to be computed within the GKS approach only when the spin currents are included in the  functional form explicitly. 
We analyze the consequence of this fact for various approximations and numerical procedures for the evaluation of SOC effects.
The practical power of the extended approach is demonstrated by  calculating the spin-orbit induced/enhanced band-splittings of inversion-asymmetric single-layer  MoSe$_2$ and inversion-symmetric bulk $\alpha$-MoTe$_2$. Key to these results is the capacity to account for SOC self-consistently while employing energy functionals and effective potentials that depend (implicitly or explicitly) on spin currents.
\end{abstract}


\maketitle

\section{Introduction}
\label{Sec1} 

Since the early days of quantum mechanics, spin-orbit interactions have played a central role in our understanding of the electronic properties of atoms, molecules, and solids. The Dirac equation and its simplified version,\cite{greiner2012relativistic} the two-component Pauli equation, were pinnacle achievements of that era, leading to a unified description of fine structure and Zeeman splittings in practically all systems known at the time, including those which would later turn out to be
 crucial to the semiconductor revolution (e.g., Ge, Si, and GaAs). \cite{yu2010fundamentals}

In this century, the discovery of nontrivial topological properties of the band structure of periodic solids, such as topological insulators and Weyl semimetals, whose extraordinary properties include quantized  transport coefficients, magnetoelectric response, chiral anomalies, non-reciprocity, etc.  has led to an explosion of interest in spin-orbit interactions.  Indeed, by making the electronic wave functions  complex even in the absence of a magnetic field, the spin-orbit interaction sets the scenario for the nontrivial response properties, band inversions and topological quantum numbers that underlie the above mentioned effects. \cite{bernevig2013topological,vanderbilt2018berry}   In a parallel development, the emergence of spintronics has raised the interest in non-collinear spin textures both in real and in momentum space.\cite{RevModPhys.76.323}  For instance, the phenomenon of spin-momentum locking -- the emergence of a spin texture in momentum space  -- is responsible for remarkable magneto-transport effects, such as the unidirectional magnetoresistance.\cite{ZG16}

In this context, it has become more pressing than ever to develop the tools of computational electronic structure so that they can be trusted to quantitatively predict the impact of spin-orbit interactions on properties such as spin-orbit splitting of bands, closing and re-openings of gaps at topological phase transitions, the  positions of conical intersection (Dirac and Weyl points) in the Brillouin zone, the dispersion of Fermi arcs and the shape of non-collinear spin textures.

Given the dominance of density functional theory (DFT) on the landscape of computational electronic structure, it seems natural to seek to include spin-orbit interaction effects through an extension of the  DFT framework.   What we mean by this is much more than simply including the spin-orbit interaction  as an additional one-body term in the Kohn-Sham equation of DFT -- an option that is already incorporated and widely available in existing electronic structure packages. Rather, in order to achieve quantitative accuracy and predictive power, we believe it is essential to include the effect of the spin-orbit interaction in the {\it many-body potentials} that appear in a (suitably generalized) Kohn-Sham theory.  

The formal framework 
for implementing this program has been known for a long time: it is the U(1) $\times$ SU(2)-invariant Spin Current DFT (SCDFT), see Refs. \onlinecite{VignaleRasolt:88, Bencheikh:03}.
This theory includes 16 external fields coupling to 16 densities, i.e., the scalar potential coupling to the particle density, the Zeeman magnetic field (three components) coupling only to the spin density, the charge vector potential (3 components) coupling only to the orbital current density, and, lastly, the SU(2) vector potential (a $3 \times 3$ tensor) coupling to the spin current densities. Depending on which functional form is invoked in the calculations, there are several convenient ways of organizing this extended set of densities and potentials; see for example Refs. \onlinecite{Goerling2006,HeatonBurgess2007,AbedinpourTokatly:10}.
Because it deals  in a unified fashion with
the magnetic interactions and the spin-orbit coupling,  SCDFT appears to be the ideal framework to simulate a  multitude of materials useful for magnetism, spintronics, orbitronics, valleytronics, and topologically non-trivial states as described above. 

In spite of its great promise, SCDFT has so far lagged behind  other DFT and non-DFT methods in its application to real material. The reason for this delay can be traced to the lack of good and  transferable approximations for the exchange-correlation (xc) energy functional in terms of spin-current densities.   
In the last two decades, it has become increasingly evident that accurate calculations of the electronic structure, including in particular band gaps and band splittings require functionals that depend on the densities not only explicitly (as in the traditional formulation of SCDFT) but also implicitly, through single-particle spin orbitals.\cite{Kurth2006,Pittalis2006,Goerling2006,HeatonBurgess2007,Sharma2007a,Sharma2007b,Goerling2018,Ullrich2018,Pluhar2019}   
The emergence of orbital-dependent functionals began with the widespread practice of including exact exchange, or a fraction thereof,  in the energy functional (the so-called ``hybrid" functionals), and gained momentum with the development of ``meta-GGA" functionals, in which the traditional set of densities is augmented by the inclusion of the (spin-)kinetic-energy density.    Most importantly for SCDFT, it was realized that spin-orbital-dependent functionals are explicitly required in any non-trivial gauge-invariant formulation.~\cite{Pittalis2017}

The problem with orbital functionals is that, because they are regarded as implicit nonlocal functionals of the density, they must be differentiated with respect to the densities in order to yield the Kohn-Sham potentials. This differentiation is difficult, as it involves the  functional derivative of the orbitals with respect to the densities.  The procedure is usually referred to as 
the ``Effective Potential Method'' (OPM),  and, while
the resulting  ``Optimized Effective Potential" (OEP) is a legitimate  local Kohn-Sham potential, the benefits of locality are  
wiped out  by the complexity and costliness of the numerical treatment.~\cite{QIN2003,Gidopoulos2012,Eich2014,Gidopoulos2013,Staroverov2023}

Experience in regular (Spin-)DFT has  demonstrated that the cost of  implementing spin-orbital functionals can be lowered, and the numerical treatment simplified by switching to a {\em generalized} Kohn-Sham (GKS) framework,~\cite{becke1993new,Seidl1996,perdew2017understanding} 
which admits the use of {\em non-local} effective potentials, as naturally appear in hybrid functional forms.
In fact, the key  ideas of this approach are also used in calculations involving meta-GGA functionals. \cite{Neumann1996,Eich2014,Lehtola2020}
In this method, as in the OPM,  the xc potential is expressed as the sum of two parts:   a functional of the spin-orbitals -- typically, but not necessarily, a fraction of the exact exchange --  plus a regular explicit functional of local densities. This simple shift in perspective   has far-reaching consequences.  The functional derivative of the explicitly orbital-dependent part of the functional yields a nonlocal, but simple potential -- the {\em Fock potential} in the case of exact exchange -- while the functional derivative of the regular part yields a local potential as in the standard Kohn-Sham formalism.  The resulting GKS equation combines the accuracy of exact nonlocal exchange with the flexibility of semilocal density functional  approximations  for the correlation energy.   Crucially, band gaps calculated in this manner become more closely related to the KS gaps, \cite{perdew2017understanding,Wing2021}since part of the derivative discontinuity of the exact functional is captured by the discontinuous dependence of the orbitals on band occupation.  The rigorous  theoretical foundation of the GKS  approach is presented in Ref.~\onlinecite{Seidl1996} for DFT. Here, we extend this approach to SCDFT.

In doing so, we lay down the framework that allows us to merge two previous works from some of the same authors of the present work:
the implementation~\cite{Desmarais2020a,Desmarais2020b,desmarais2021spin2,desmarais2019spinI} and application~\cite{Comaskey2022,Bodo2022} of the (regular) global hybrids in SCDFT via the \textsc{Crystal} code 
and the (formal) proposal of Meta-GGAs for SCDFT.~\cite{Pittalis2017}
This allows us to demonstrate that at the heart of the success of the method is its ability to include the dependence of the effective many-body potentials on spin currents. Even when this dependence is only implicit (as discussed below) its inclusion is essential to obtain agreement with experimental results --- whereas conventional treatments of spin-orbit coupling fail. When the inclusion of spin currents is explicit, it  becomes necessary to include  the feedback on the effective-vector potentials explicitly as well.

The paper is organized as follows: We start with an introductory section on the (regular) Kohn-Sham approach to SCDFT. We then describe the GKS approach and proceed to its application. 
We discuss first the prominent case of exact exchange  followed by an in-depth discussion of global hybrid forms. 
We also discuss the case of spin-current dependent  Meta-GGAs. In this way, we are able to highlight several crucial features which are peculiar to the GKS approach of SCDFT.
We apply the approach  to the calculations of  valence-band splittings induced/enhanced by SOC in inversion-asymmetric single-layer, 2D, MoSe$_2$ with spin-splitting (Rashba-I effect) and inversion-symmetric bulk $\alpha$-MoTe$_2$ with spin-valley locking (Rashba-II effect). 
After touching on  near-future developments, we conclude.

\section{Formal Aspects}

\subsection{Spin-Current Density Functional Theory}
\label{Sec2}

In order to appreciate the key difference between SCDFT and Spin-DFT (SDFT, the most popular flavor of DFT), it is useful to start with the SDFT Hamiltonian:
\begin{widetext}
\begin{eqnarray}\label{H_SDFT}
\hat{H}_\SDFT =  
 \frac{1}{2} \int d^3r~ \hat{\Psi}^\dagger({\bf r}) \left(-i \nabla\right)^2
\hat{\Psi}({\bf r})  +  \int d^3r~ \left[ \hat{n}({\bf r}) v({\bf r}) + \hat{m}^a ({\bf r}) {B}^a({\bf r}) \right] + \hW \;,
\end{eqnarray}
\end{widetext}
where $\FPdr = (\FPdur,\FPddnr)$  denotes a  two-component creation field operator ($\uparrow$ and $\downarrow$ refer to  the spin ``up'' and ``down''); $v$ represent an external scalar-multiplicative  potential
that couples to the electrons via the particle-density operator
$\hat{n} =  \hat{ \Psi}^{\dagger}  \hat{ \Psi}$; ${B}^a$ is the $a$ compoenent of a magnetic field that couples to the electrons via the spin-density operator $\hat{m}^a= \hat{ \Psi}^{\dagger}  {\sigma}^a \hat{ \Psi}$ (${\sigma}^a$ denotes the Pauli matrices $\sigma^x, \sigma^y, \sigma^z$).
The last term of the right hand side denotes the electron-electron interaction
\begin{align}\label{W}
\hW &=   \int d^3r \int d^3r' \frac{\FPd(\br) \FPd(\br^\prime)  \FP(\br^\prime) \FP(\br)}{2|\br - \br^\prime|}\;.
\end{align}
Unless otherwise stated, we use Hartree atomic units in which $\hbar = m = 1$. 
Also note that, for notational convenience, the  Bohr magneton $\mu_B$ factor is absorbed in the symbol ${B}^{a}$.

The SDFT Hamiltonian does {\em not} include either the vector potential corresponding to ${B}^a$ nor spin-orbit couplings. For taking into account the latter interactions in DFT fashhion, 
we  consider the SCDFT Hamiltonian, which can be obtained via the 
``minimal'' substitution $-i \nabla \rightarrow -i \nabla + \frac{1}{c} {\bf A}({\bf r})  +  \frac{1}{c}    \sigma^{a}  {\bf A}^{a}({\bf r})$:
\begin{widetext}
\begin{eqnarray}\label{H_1}
\hat{H}_\SCDFT =
\frac{1}{2} \int d^3r~ \hat{\Psi}^\dagger({\bf r}) \left[  -i \nabla + \frac{1}{c} {\bf A}({\bf r})  +  \frac{1}{c}    \sigma^{a}  {\bf A}^{a}({\bf r}) \right]^2 
\hat{\Psi}({\bf r}) + \int d^3r~ \left[ \hat{n}({\bf r}) v({\bf r}) + \hat{m}^a ({\bf r}) {B}^a({\bf r}) \right] + \hW \;.
\end{eqnarray}
\end{widetext}
Above and in the following, we denote with bold characters, $\boldsymbol{A}$, quantities with spatial indices (Greek lower indices, $A_\mu$, when written explicitly); and use an arrow, $\vec{A}$, to denote quantities with spin indices (upper Latin indices, $A^a$, when written explicitly). 
Thus $\vbA$ denotes a tensor with two indices $A^a_\mu$. Both $\mu$ and $a$ have values $x$,$y$,$z$. Contractions over spatial indices are denoted with ``$\cdot$'', Einstein convention is applied to intend summation over repeated indices.

Eq. \eqref{H_1} includes besides the terms of the SDFT Hamiltonian also a 
(charge-) vector potential $\bA(\br)$ and a spin-vector potential $\bA^a(\br)$. 
While $\bA(\br)$ is useful to represent an external magnetic field in the usual fashion $B^a = \epsilon^{a \mu \nu} \partial_\mu A_\nu$,
 $\bA^a(\br)$ is useful to represent the (one-body) spin-orbit couplings in the system.~\cite{FroehlichStuder:93,Bencheikh:03,AbedinpourTokatly:10}
Note, in our notation we absorb the prefractor $\frac{\mu_B}{2}$ in the symbol ${\bf A}^a$.
These vector potentials may be viewed as some ``induction'' fields, in the sense that they induce 
particle and spin currents in the systems on which they act. 
To make the latter fact apparent and towards a proper density functionalization, let us expand the first term in Eq. \eqref{H_1}
\begin{widetext}
\begin{eqnarray}\label{H-SCDFT}
\hat{H}_\SCDFT = \hT + \hW  +  \int d^3r~ \hat{n}({\bf r})\tilde{v}({\bf r})
+  \int d^3r~  \hat{m}^a({\bf r})  \tilde{B}^a({\bf r}) + \frac{1}{c} \int d^3r~ {\hat {\bf j}}(\br)  \cdot {\bf A}(\br) + \frac{1}{c} \int d^3r~ {\hat { \bf J}^a}(\br) \cdot {{\bf A}^a }(\br)\;,
\end{eqnarray}
\end{widetext}
where
\begin{align}\label{T}
  \hT = \int d^3r \; \FPd(\br) \left( -\frac{\nabla^2 }{2} \right) \FP(\br)\;,
\end{align}
\begin{equation}
\tilde{v} =   v + \frac{1}{2c^2 }  \left[ {\bf A} \cdot {\bf A} + {\bf A}^a \cdot {\bf A}^a \right]\; ,
\end{equation}
and
\begin{equation}
\tilde{B}^a =  B^a +  \frac{1}{2 c^2}  {\bf A} \cdot {\bf A}^a\;.
\end{equation}

Crucially, in Eq. \eqref{H-SCDFT} we also find the  (paramagnetic) particle-current operator 
$
\hat{{\bf j}}  = 
\frac{1}{2i} \left[ \hat{ \Psi}^{\dagger}  \nabla \hat{ \Psi}  - \left( \nabla \hat{ \Psi}^{\dagger} \right) \hat{ \Psi}\right]
$
and the  (paramagnetic) spin-current operator
$
\hat{{\bf J}}^a = 
\frac{1}{2i} \left[ \hat{ \Psi}^{\dagger}  {\sigma}^a \nabla \hat{ \Psi}  -
\left( \nabla \hat{ \Psi}^{\dagger} \right) {\sigma}^a \hat{ \Psi}\right]\;.
$ 
Therefore we may anticipate that while SDFT may only accounts for particle- and spin-density self-consistently, 
SCDFT must account also for particle- and spin-current self-consistently (below).

In fact, given the external fields $v$,  ${B}^a$, ${\bf A}$, and ${\bf A}^a$, the ground-state energy may then be determined by means of a constrained-search minimization:~\cite{Levy:82,Lieb:83}
\begin{widetext}
\begin{align}\label{Emin1}
E = \min_{(n,~ \vm,~ {\bf j},~ \vbJ)} \Big\{ F[n, \vm, {\bf j}, \vbJ ]  +   \int d^3r~ {n}({\bf r})\tilde{v}({\bf r})
+  \int d^3r~  {m}^a({\bf r})  \tilde{B}^a({\bf r}) +  \frac{1}{c} \int d^3r~ {{\bf j}}(\br)  \cdot {\bf A}(\br) +  \frac{1}{c} \int d^3r~ {{ \bf J}^a}(\br) \cdot {{\bf A}^a }(\br) 
\Big\} \; ,
\end{align}
\end{widetext}
with
\begin{align}\label{F}
F[n,  \vm, {\bf j}, \vbJ ] = \min_{ \Psi \rightarrow (n,~ \vm,~ {\bf j},~ \vbJ)} \langle \Psi | \hat{T} + \hat{W} | \Psi \rangle \;.
\end{align}
In Eq. \eqref{Emin1}, the inner minimization stated in
Eq. \eqref{F}
is carried out over all the  antisymmetric many-electron  wave functions yielding the
prescribed  set of densities and the outer minimization is carried out with respect to all $N$-representable densities (see note at Ref. \onlinecite{note6} for more details). Eq. \eqref{F} defines a universal functional of the densities and currents. The term ``universal'' (as usual) highlights the fact that its definition does not involve external potentials. 

The Kohn-Sham scheme in SCDFT invokes the non-interacting universal functional:
\begin{align}\label{Ts}
T_\KS[n,  \vm, {\bf j}, \vbJ  ] = \min_{ \Phi \rightarrow (n,  \vm, {\bf j}, \vbJ)} \langle \Phi | \hat{T}  | \Phi \rangle \;,
\end{align}
which is obtained from Eq. \eqref{F} by setting $\hW=0$. Here and in the following $\Phi$ denotes a Slater determinant of $N$ single-particle orbitals as opposed to more general $N$-particle antisymmetric wave functions $\Psi$.  Crucially, assuming that the same set of densities is both interacting and non-interacting $v$-representable, one may further decompose $F$ as follows:
\begin{equation}
\label{FKS}
F[n,  \vm, {\bf j}, \vbJ] = T_\KS[n,  \vm, {\bf j}, \vbJ] + E_H[n] + E_{\rm xc}[n,  \vm, {\bf j}, \vbJ  ] \; , 
\end{equation}
in terms of the KS kinetic energy 
$T_\KS[n,  \vm, {\bf j}, \vbJ  ]$, the Hartree energy $E_{\rm H}[n] = \frac{1}{2} \iint ~ \frac{{n}({\bf r}) {n}({\bf r}')}{|\br - \br'| }$ and a
remainder, $E_{\rm xc}[n,  \vm, {\bf j}, \vbJ] $ --- the xc-energy functional in SCDFT. 

Given $E_{\rm xc}$, or an approximation thereof in practice,
the  problem of determining the ground-state energies of an interacting system is therefore translated into finding the ground state of a non-interacting system. 
The KS equations in SCDFT have the form of single-particle Pauli equations including scalar, vector, and magnetic fields:~\cite{VignaleRasolt:88,Bencheikh:03}
\begin{equation}\label{KSeq}
\left[ \frac{1}{2}\left( -i   \nabla + \frac{1}{c} { \mathbfcal{A} }_\KS  \right)^2 +  {\cal V}_\KS  \right] \Phi_k = \varepsilon_k \Phi_k \; ,
\end{equation}
where
\begin{align}\label{calA}
{ \mathbfcal{A} }_\KS &=  
\left( {\bf A} + \sigma^a {\bf A}^a \right) +   \left(  {\bf A}_{{\rm xc}} + {\bf A}^a_{{\rm xc}} \right)
\nonumber \\
&={ \mathbfcal{A} } + { \mathbfcal{A} }_{\rm xc}\;,
\end{align}
\begin{align}
{\cal V}_\KS &= v_{\rm H} + \left( v + v_{\rm xc} \right) + \sigma^a\left( B^a + B^a_{\rm xc} \right) \nonumber \\
&+ \frac{1}{2c^2} \left[ \mathbfcal{A}^2-  {\mathbfcal A}^2_\KS \right]\;,
\end{align}
in which
\begin{equation}\label{aAxc}
 {\frac{1}{c}}{\bf A}_{\rm xc}(\br) = \frac{\delta E_{\rm xc}}{\delta {\bf j}(\br)}   
\end{equation}
is an Abelian xc-vector potential,
\begin{equation}\label{naAxc}
{\frac{1}{c}}{\bf A}^a_{\rm xc}(\br) = \frac{\delta E_{\rm xc}}{\delta { \bf J}^a(\br)}
\end{equation}
is the $a$-th component of a non-Abelian xc-vector potential,
$
{B}^a_{\rm xc}(\br) = \frac{\delta E_{\rm xc}}{\delta {m^a}(\br)}
$
is the $a$-th component of a xc-magnetic potential,
$
 v_{\rm xc}(\br) = \frac{\delta E_{\rm xc}}{\delta n(\br)}
$
is a xc-scalar potential, and $v_H(\br) = \int d\br \frac{n(\br')}{|\br - \br'|}$ is the usual Hartree potential.
The KS densities are
obtained from the (occupied) KS spinors as follows:
\begin{align}\label{Gamma-n}
n_\KS({\bf r}) & = \sum_{k=1}^{N} \Phi^\dagger_k({\bf r}) \Phi_k({\bf r}) \;,
\end{align}
already used  in the expression of $v_{\rm H} $,
\begin{align}\label{Gamma-s}
{\vec m}_\KS({\bf r}) &= \sum_{k=1}^{N} \Phi^\dagger_k({\bf r})\;{\vec \sigma}\; \Phi_k({\bf r}) \; ,
\end{align}
\begin{equation}\label{Gamma-j}
{\bf j}_\KS({\bf r}) = \frac{1}{2i} \sum_{k=1}^{N} \Phi^\dagger_k({\bf r}) \left[ \nabla \Phi_k({\bf r}) \right] -  \left[ \nabla \Phi^\dagger_k({\bf r}) \right] \Phi_k({\bf r}) \; ,
\end{equation}
and
\begin{align}\label{Gamma-J}
{\vec {\bf J}}_\KS({\bf r}) &= \frac{1}{2i} \sum_{k=1}^{N} \Phi^\dagger_k({\bf r}) {\vec \sigma}  \left[ \nabla \Phi_k({\bf r}) \right] -  \left[ \nabla \Phi^\dagger_k({\bf r}) \right] {\vec \sigma} \Phi_k({\bf r})\;.
\end{align}
By virtue of the non-interacting $v$-representability assumption, the {\em exact} $E_{\rm xc}$ yields the exact interacting densities, which coincide with the KS densities: $n_\KS \equiv n$, ${\vec m}_\KS \equiv {\vec m}$,  ${\bf j}_\KS  \equiv {\bf j}$, and  $\vbJ_\KS \equiv \vbJ$.\\

As argued in the Introduction, spin-orbital functionals can enable sufficiently general SCDFT applications.
For determining the effective {\em local} potentials from spin-orbital dependent functionals, however,
an extra set of integro-differential equations needs then to be solved for determining the corresponding {\em local}  potentials. Such a
numerical task is subtle,~\cite{QIN2003,Gidopoulos2012,Eich2014,Gidopoulos2013,Staroverov2023} and it usually  exceeds  the  cost of  more straightforward generalized-gradient-approximations (GGA).
Fortunately, the cost involved in the application of spin-orbital functionals can be lowered, and the corresponding numerical implementations can also be simplified, by invoking an appropriate {\em exact} generalization of the KS approach. 
This is usually handled by admitting {\em partially interacting} KS systems, which exhibit {\em non-local} effective potentials.~\cite{Seidl1996,Garrick2022} Below, we spell out  and analyze the case for SCDFT.

\subsection{From Regular to Generalized-KS Systems in SCDFT}

GKS systems can be introduced in SCDFT in a way that is similar to  (S)DFT by noting that 
the minimization in Eq. \eqref{Emin1} can {\em equivalently} be performed by invoking different
splittings of $F[n,  {m}^a, {\bf j}, { \bf J}^a ]$ and a different minimization procedure. In detail, let us consider:
\begin{align}\label{GK-F}
F[n,  \vm, {\bf j}, \vbJ] = F_\GKS[n,  \vm, {\bf j}, \vbJ] + E^\GKS_{\rm Hxc}[n,  \vm, {\bf j}, \vbJ] \; ,
\end{align}
where
\begin{align}\label{FS}
F_\GKS[n,  \vm, {\bf j}, \vbJ] = \min_{ \Phi \rightarrow (n,  \vm, {\bf j}, \vbJ)} \langle \Phi | \hO_\GKS  | \Phi \rangle 
\end{align}
is the analogous of Eq. \eqref{Ts} but here
$\hO_\GKS $ may differ from $\hT$ by including some interaction (below).
Next, note that
\begin{widetext}
\begin{align}\label{EminGKS}
E &=  \min_{(n,  \vm, {\bf j}, \vbJ)}~ \Big\{ \min_{ \Phi \rightarrow (n,  \vm, {\bf j}, \vbJ)}  \langle \Phi | \hO_\GKS  | \Phi \rangle 
+ E^\GKS_{\rm Hxc}\Big[n,  \vm, {\bf j}, \vbJ\Big]  \nonumber \\
&+  \int d^3r~ {n}({\bf r})\tilde{v}({\bf r})
+  \int d^3r~  {m}^a({\bf r})  \tilde{B}^a({\bf r}) +  \frac{1}{c} \int d^3r~ {{\bf j}}(\br)  \cdot {\bf A}(\br) +  \frac{1}{c} \int d^3r~ {{ \bf J}^a}(\br) \cdot {{\bf A}^a }(\br) 
\Big\} \nonumber \\
&=  \min_{\Phi}~ \Big\{  \langle \Phi | \hO_\GKS   | \Phi \rangle 
+ E^\GKS_{\rm Hxc}\Big[n[\Phi],  {\vm}[\Phi], {\bf j}[\Phi], \vbJ[\Phi] \Big]  \nonumber \\
&+  \int d^3r~ {n}[\Phi]({\bf r})\tilde{v}({\bf r})
+  \int d^3r~  {m}^a[\Phi]({\bf r})  \tilde{B}^a({\bf r}) +  \frac{1}{c} \int d^3r~ {{\bf j}}[\Phi](\br)  \cdot {\bf A}(\br) +  \frac{1}{c} \int d^3r~ {{ \bf J}^a}[\Phi](\br) \cdot {{\bf A}^a }(\br) 
\Big\}
\end{align}
\end{widetext}
may  be admissible, provided  interacting and non-interacting $N$- and $v$-representability
hold true. 

In practice, the form of GKS schemes depends upon the detail of $\hO_\GKS$ and $E^\GKS_{\rm Hxc}$.
A prominent example is $\hO_\GKS \equiv \hT + \alpha \hW$ where $\alpha \in (0,1]$ \textit{turns on} the interaction in the GKS reference system --- yet the minimization is restricted to single Slater determinants $\Phi$, only ---
and $E^{\GKS}_{\rm Hxc} \equiv (1-\alpha) E^\DFA_{\rm Hx} + E^\DFA_{\rm c} \equiv E^\DFA_{\rm Hxc}$; i.e., 
we mix Fock exchange with (standard) LDAs or GGAs, in the form of typical global hybrid approximations. 

Approximations of this kind can fix, at least partially, the self-interaction error of DFAs. 
They have been justified by the necessity of mimicking an exact (almost) semi-local xc-hole by the combination of a non-local exact-exchange hole with an  approximate (semi-)local correlation hole. Global hybrids and refinements thereof,  have been guided by the (so-called) adiabatic connection integration.
\cite{harris1984adiabatic,becke1993new,perdew1996rationale,KochHolthausen2001,Garrick2022}
But there is no  prescription for choosing an optimal value of the hybridization parameter $\alpha$  that works for all systems. 
One needs to devise ways  to find optimal values, driven by first principle calculations,
\cite{Marques2011,skone2014self,MIOSCHYB} or consider more sophisticated forms and procedure of hybridizations.\cite{Kronik2012,Nguyen2018,Wing2021,Prokopiou2022}
Furthermore, the optimization task always targets specific observables; most commonly, energy  gaps. 
These aspects have received a huge attention and application in (S)DFT. {\em In this work we show that switching from Spin-DFT to SCDFT is crucial for determining optimal mixing that can work both for band gaps and band splittings of spin-orbit coupled materials.}

Hence, let us start with the functional form
\begin{widetext}
\begin{eqnarray}\label{FunctionalGKS}
E[\Phi] &=& T_\GKS[\Phi] + \alpha E^{\rm Fock}_{\rm x} [\Phi]  + E^\DFA_{\rm Hxc}\Big[n[\Phi],  {\vm}[\Phi], {\bf j}[\Phi], \vbJ[\Phi] \Big]  
+  \int d^3r~ {n}[\Phi]\tilde{v}({\bf r})+  \int d^3r~  v[\Phi]({\bf r})  \tilde{B}^a({\bf r}) \nonumber \\
&+&  \frac{1}{c} \int d^3r~ {{\bf j}}[\Phi](\br)  \cdot {\bf A}(\br) +  \frac{1}{c} \int d^3r~ {{ \bf J}^a}[\Phi](\br) \cdot {{\bf A}^a }(\br)\;, \nonumber
\end{eqnarray}
\end{widetext}
where
\begin{eqnarray} \label{eq:WW}
E^{\rm Fock}_{\rm x} [\Phi] \equiv - \frac{1}{2} \int d^3 r \int d^3 r'~\frac{ {\rm Tr} \left\{ \Gamma({\bf r},{\bf r}')\Gamma({\bf r}',{\bf r}) \right\} }{ |{\br - \br'}| }\;,\end{eqnarray}
is the Fock exchange, here evaluated with GKS spinors, and
\begin{align}
\label{GammaKS}
\Gamma({\bf r},{\bf r}') \equiv \sum_{k=1}^{N} \Phi_k({\bf r})\Phi^\dagger_k({\bf r}')\;
\end{align}
is the one-electron reduced density matrix (1RDM). In Eq. \eqref{eq:WW}, Tr denotes the trace over spin. 

As announced, we shall consider   ``typical'' global hybrids forms. With ``typical'', we intend  forms that mix  a fraction of Fock exchange, $E^{\rm Fock}_{x}[\Phi]$, with GGAs (or lower rung approximations).

Note, GGAs in SCDFT (as in Spin-DFT) may dependent on all the basic variables and their gradients, but not on other quantities (e.g. kinetic energy densities).

The corresponding (generalized) KS equation, then, reads as follows
\begin{eqnarray}
\label{GKS-RES}
\hat{H}_{\GKS} &=& \frac{1}{2}\left( -i  \nabla + \frac{1}{c} {\mathbfcal{A}}_\GKS \right)^2 +  \alpha \hat{{\cal V}}_{\rm x}^{\rm NL}   \nonumber \\
&+& { \cal V }_{\GKS} \;
\end{eqnarray}
where
\begin{eqnarray}
\label{GKS-A}
{ \mathbfcal{A}_\GKS } =  \mathbfcal{A} +  (1 -\alpha) {\mathbfcal{A}}^{\DFA}_{\rm x} + {\mathbfcal{A}}^{\DFA}_{\rm c}
 \end{eqnarray}
 with
\begin{align}
\mathbfcal{A}^\DFA_{\rm x/c} =  {\bf A}^\DFA_{\rm x/c} + \sigma^a {\bf A}^{\DFA,a}_{\rm x/c}\;,
\end{align}
\begin{align}
\label{eq:VNL}
\hat{{\cal V}}^{\rm NL}_{\rm x}  \Phi_k  &= \frac{\delta E^{\rm Fock}_{\rm x} [\Phi]}{\delta{\Phi_k}^{\dagger}} =
-
\int d^3r'~   \frac{\Gamma({\bf r},{\bf r}') \Phi_k({\bf r}') }{|{\bf r} - {\bf r}'|}\;;
\end{align}
$\hat{{\cal V}}_{\rm x}^{\rm NL}$ is the Non-Local Fock potential ---  here evaluated  with GKS spinors.
Next,
 \begin{eqnarray}
\label{GKS-V}
{ \cal{V}_\GKS } &=& {\cal{V}} +  v_{\rm H} + (1 -\alpha) {\cal{V}}^{\DFA}_{\rm x} + {\cal{V}}^{\DFA}_{\rm c}
\nonumber \\
&+&  \frac{1}{2c^2} \left[ \mathbfcal{A}^2-  {\mathbfcal A}^2_\GKS \right]
 \end{eqnarray}
with
\begin{align}
v_{\rm H}  &=  \int d^3r'~ \frac{{\rm Tr}\;  \Gamma ({\bf r}',{\bf r}') }{|{\bf r} - {\bf r}'|}\;
\end{align}
and
\begin{align}
\label{VDFA}
{\cal V}^\DFA_{\rm x/c} =   \left( v^{\DFA}_{\rm x/c} + \sigma^a B^{\DFA,a}_{\rm x/c} \right)\;.
\end{align}

The GKS equations reduce to the regular KS equations for $\alpha=0$.
For $\alpha \ne 0$, the xc-scalar, xc-magnetic, and xc-vector potentials produced by $E^\DFA_{\rm Hxc}$ get, partially, 
replaced by a fraction of the non-local potential $\hat{{\cal V}}_{\rm x}^{\rm NL}$. At $\alpha = 1$, the DFA gives no contribution: we end up with the HF approximation for the Pauli equation. 
In passing, also note that,  had we invoked a meta-GGA instead of a GGA, the differentiation w.r.t. the spin-orbitals would have generated additional terms from the explicit dependence on the (spin-)kinetic-energy density --- yielding to terms like the ones already accounted for in (non-collinear) SDFT.\cite{Peralta2007}

It is expedient to contrast the GKS equations including exact-exchange, Eqs. \eqref{GKS-RES}-\eqref{VDFA}, against the exact-exchange approximation of the regular KS approach. In the present GKS scheme, the non-local Fock potential is directly given in terms of
$\Gamma({\bf r},{\bf r}')$ [see Eq. \eqref{eq:VNL}]. On the other hand, in the regular KS approach to SCDFT, exact-exchange leads to the 16 integro-differential OEP equations that produce 16 local exchange potentials 
in response to variations in 16 basic density components.\cite{Goerling2006,HeatonBurgess2007} 
At the present stage of development, 
the determination of  local exact-exchange potentials are both numerically more involved and more costly than the evaluation of the non-local Fock potential.

\section{ SCDFT {\em versus} SDFT+SOC }
\label{J-Fock}

The GKS-SCDFT framework allows us to include SOC non-perturbatively {\em and} self-consistently in a density functional calculation. It is interesting to see how this general framework adapts to various approximations.  For sake of simplicity, let us restrict to systems with vanishing magnetization ($\vm = 0$) and vanishing particle currents ($\bj=0$) but with non-vanishing spin currents ($\vbJ \ne 0$, i.e. a typical time-reversal symmetry preserving system with SOC). Additionally, we consider systems for which 
spin-currents are vanishing when SOC is turned off.
Thus, there are two main types of approximations to be considered: approximations depending on all the basic densities 
but currents; and those also including currents. Of the latter class, there is also the case of those approximations that include
currents but only implicitly. In view of the complexity of this scenario, and in preparation of the calculations we shall perform in the next section,
we first review analogous  {\em Gedanken} calculations.

{\em Standard GGAs}.~ As a first example, we may perform a calculation by using a  GGA that, like any standard GGA, does {\em not} include a dependence on spin currents. In detail,
\begin{eqnarray}\label{GGA}
 E^{\rm GKS}_{\rm xc}[\Phi] \approx  E^{\rm GGA}_{\rm xc}[n[\Phi]]\;.
 \end{eqnarray}
Correspondingly, Eq. \eqref{aAxc} implies $\vbA^{\rm GGA}_{\rm xc} = 0$.
Then, the GKS equations we must solve look like the GKS equations of SDFT calculations, if it were not that  SOC is also added directly.
Therefore,  more appropriately, we should  
regard the resulting equations as some {\em approximate} GKS SCDFT equations. This interpretation is forced on us by the fact that, by construction, the SDFT energy functional is derived for systems described by the Hamiltonian reported  in Eq. \eqref{H_SDFT}: i.e., for systems that
do {\em not} include SOC. Via SDFT we may, however, include SOC as a {\em perturbation}. 
This state of affair must be borne in mind also when
discussing the more involved cases reported below.
An evaluation of SOC can be performed in a ``second variational'' step after the convergence of a ``first variational'' SDFT calculations.\cite{Huhn2017,koelling1977technique} Here, the ``first variational'' step is the diagonalization of the SDFT Hamiltonian (without SOC) in a set of basis functions, and the ``second variational'' step is the diagonalization of the SDFT Hamiltonian plus SOC in the basis of SDFT single particle orbitals (a one-shot calculation, that is ordinary first-order quasi-degenerate perturbation theory).
It may be tempting to iterate the procedure self-consistently until convergence but, at the level of a regular GGA, the effects of self-consistency are, usually, irrelevant.\cite{desmarais2021perturbation,desmarais2023perturbation}
Our formalism makes it apparent that the degree of the aforementioned self-consistency {\em cannot} make up for the missing dependence of an xc-approximation on the spin currents in the selected GGA.

{\em Global Hybrid}.~ As a more subtle and advanced example,  let us perform a calculation by employing  a {\em regular} hybrid functional; i.e. consisting in a fraction of Fock exchange plus a complementary fraction of a regular GGA. In detail,
\begin{widetext}
\begin{eqnarray}\label{rDFA}
 E^{\rm GKS}_{\rm xc}[\Phi] \approx  E^{\rm Hybrid}_{\rm xc}[\Phi]  \equiv  \alpha E^{\rm Fock}_{\rm x} [\Phi]  + (1-\alpha)E^{\rm GGA}_{\rm x}[n[\Phi]] +  E^{\rm GGA}_{\rm c}[n[\Phi]]\;.
 \end{eqnarray}
\end{widetext}

In the absence of SOC, the Fock exchange can be restricted  to the usual one component (i.e., globally collinear) form. When SOC is included, the Fock exchange must be  upgraded to a two-component (i.e. non collinear) form --- as we have described in the previous sections. 
Next, to appreciate the difference of more or less self-consistent calculations, it is instrumental to scrutinize  the short-range behaviour of the Fock-energy density.\cite{Pittalis2017}

For this purpose, coming back to Eq. \eqref{eq:WW}, let us employ the shorthand notation
\begin{equation}\label{eq:Qx}
Q_{\rm x}({\bf r},{\bf r}') = {\rm Tr} \left\{ \Gamma({\bf r},{\bf r}')\Gamma({\bf r}',{\bf r}) \right\} \; ,
\end{equation}
and change integration variables by introducing the inter-particle separation $\mathbf{u}$: 
\begin{eqnarray} \label{Ex-Qx}
E^{\rm Fock}_{\rm x}[\Phi] &=& - \frac{1}{2} \int d^3 r \int d^3 u~\frac{ Q_{\rm x}({\bf r} + {\bf u}/2,{\bf r} - {\bf u}/2) }{ u}\;. \nonumber
\end{eqnarray}
We recall that $Q_{\rm x}$ from Eq. \eqref{eq:Qx} is the trace of a $2 \times 2$ matrix, and thus may be decomposed in the basis $I$, $\sigma^x, \sigma^y, \sigma^z$. Next, a Taylor expansion of the spherical average around $\mathbf{r}$, $\langle Q_{\rm x} \rangle$,  to second-order in $u$ gives:
\begin{widetext}
\begin{eqnarray}\label{T-Qx}
\langle Q_{\rm x}  ({\bf r}, u ) \rangle  =  \frac{ n^2(\br) }{2} +  \frac{u^2}{6} 
\left[ 2n(\br) \tau(\br)     -    \vbJ(\br) \odot \vbJ(\br)  - \frac{  n(\br)\; \nabla^2 n(\br)}{4}   \right] + \mathcal{O} \left( u^4 \right)
 \;.
\end{eqnarray}
\end{widetext}
Here, ``$\circ$'' denotes contraction w.r.t. spin indices, and ``$\odot$'' denotes a double contraction w.r.t. spin and real-space indices,
\begin{equation}
\tau({\bf r}) = \frac{1}{2} \sum_{k=1}^{N} \Big( \nabla \Phi_k^{\dagger}({\bf r}) \Big) \cdot \Big( \nabla \Phi_k({\bf r}) \Big)\;,
\end{equation}
is the  kinetic energy density of the occupied GKS spinors.
Eq. \eqref{T-Qx}  shows that both the particle density and the spin currents  contribute --- via the spinors --- to the Fock exchange energy-density at short range and, thus, they also contribute to the corresponding non-local potential $\hat{{\cal V}}_{\rm x}^{\rm NL}$ [see Eq. \eqref{eq:VNL}]
In passing also note that because the kinetic energy density does {\em not} belong to the basic variables of SCDFT, it must be regraded as a purely spinor-dependent term.

From Eq. \eqref{T-Qx}, we can appreciate the important fact that a {\em perturbative} evaluation of SOC done in GKS-SDFT at the level of a hybrid approximation unavoidably {\em misses} the feedback from the implicit dependence on the spin currents. For going beyond such a perturbative evaluation, it is necessary to include SOC self-consistently. Adding SOC will turn on the spin-currents. Therefore,
unconstrained two-component spin-orbital {\em non}-perturbative GKS-SDFT+SOC calculations are nothing else  but GKS-SCDFT calculations.
 Because the  dependence of $E^{\rm Fock}_{\rm x}[\Phi]$ on the spin currents is implicit, however,
such an equivalence remains subtle.
The crucial point is that, at variance with the previous {\em GGA-only} case, 
the self-consistency in this latter case gains on the feedback from the spin currents:
both via the energy functional $E^{\rm Fock}_{\rm x}[\Phi]$ and in the non-local potential $\hat{{\cal V}}_{\rm x}^{\rm NL}$.
The effect of such a self-consistency may {\em not} be expected to be irrelevant.

{\em Spin-current-dependent MGGAs}.~Let us lastly perform a calculation that invokes a functional form with an {\em explicit}  dependence on the spin currents.
This type of form  can be generated, for example, by ``localizing'' the Fock exchange functional.\cite{Pittalis2017} A most direct way of doing this is to perform a Gaussian re-summation of Eq. \eqref{T-Qx}. 
Straightforwardly, one gets
\begin{equation}
\label{JLP}
E^{\rm JLP}_{\rm x}[\Phi]  = - \frac{3\pi}{4} \int d^3r~ \frac{n^3(\br)}{ \left[  \tau(\br)   - \frac{ \vbJ(\br) \odot \vbJ(\br)  }{ 2n(\br) } - \frac{ \nabla^2 n(\br)}{8}   \right]} \;,
\end{equation}
Eq. \eqref{JLP} is a spin-current dependent generalization of the Lee-Parr functional for exchange (JLP),\cite{Pittalis2017,LP87} which, through Eq. \eqref{naAxc}, gives rise to the non-Abelian x-vector potential 
\begin{equation}
\label{AJLP}
\vbA^{\rm JLP}_{\rm x}  = - \frac{3\pi c}{16}  \frac{n^2(\br)  }{\left[  \tau(\br)     
- \frac{ \vbJ(\br) \odot \vbJ(\br)  }{ 2 n(\br) }  - \frac{ \nabla^2 n(\br)}{8} \right]^2} \vbJ(\br) \;.
\end{equation}

{\em Therefore, Eq. \eqref{AJLP} makes the switch from SDFT+SOC to SCDFT explicit also at the level of the GKS equations.}
The contribution of $\vbA^{\rm JLP}_{\rm x}$ may not be expected to be minor.

In the next section, we shall make use of JLP to {\em reduce} the cost of the numerical calculation
of hybrid calculation by using
\begin{widetext}
\begin{equation}\label{H@JLP}
 E^{\rm GKS}_{\rm xc}[\Phi] \approx E^{\rm JMGGA}_{\rm xc}[\Phi] \equiv \alpha E^{\rm JLP}_{\rm x}[\Phi] + (1- \alpha) E^{\rm GGA}_{\rm x}[n] + E^{\rm GGA}_{\rm c}[n]\;,
\end{equation}
\end{widetext}
where $E^{\rm Fock}_{\rm x}[\Phi]$ in Eq. \eqref{rDFA} has been replaced with $E^{\rm JLP}_{\rm x}[\Phi]$.

In passing, we recall that the importance of the dependence of density functionals  on the  {\em particle} current has been largely demonstrated.\cite{Dobson93,Becke96,Becke-j02,J-PBE,burnus2005time,TP05,Pittalis07,Pittalis09,Rasanen09,Oliveira2010,Tricky16,Furness2016,Holzer2022}
The point stressed above concerns instead the {\em spin} current which, to the best of our knowledge, remains to be explored.

Last but not least, the derivation  just above may be extended to generate  MGGA-correlation forms by reducing the {\em non}-locality of higher-level xc models. This is left to future efforts together with other appealing possibilities as outlined in the road map of developments given below.

\section{Applications}
\label{Applications}

The aim of this section is to demonstrate the practical importance and flexibility of the GKS approach of SCDFT for going beyond the perturbative treatment of SOC.
We shall consider the  case of the SOC-induced/enhanced band splittings that occur near the top of the valence band of  layered molibdenum dichalcogenides. These systems are
{\em time-reversal symmetric} and the ground states have vanishing  magnetization ($\vm=0$), vanishing particle current ($\bj=0$),
and, because of the presence of SOC, non-vanishing spin currents ($\vbJ \ne 0$). The calculations  we consider here showcase numerically the different SOC evaluations discussed formally in Sec. \ref{J-Fock}.

\begin{figure}[t!]
\centering
\includegraphics[width=8.6cm]{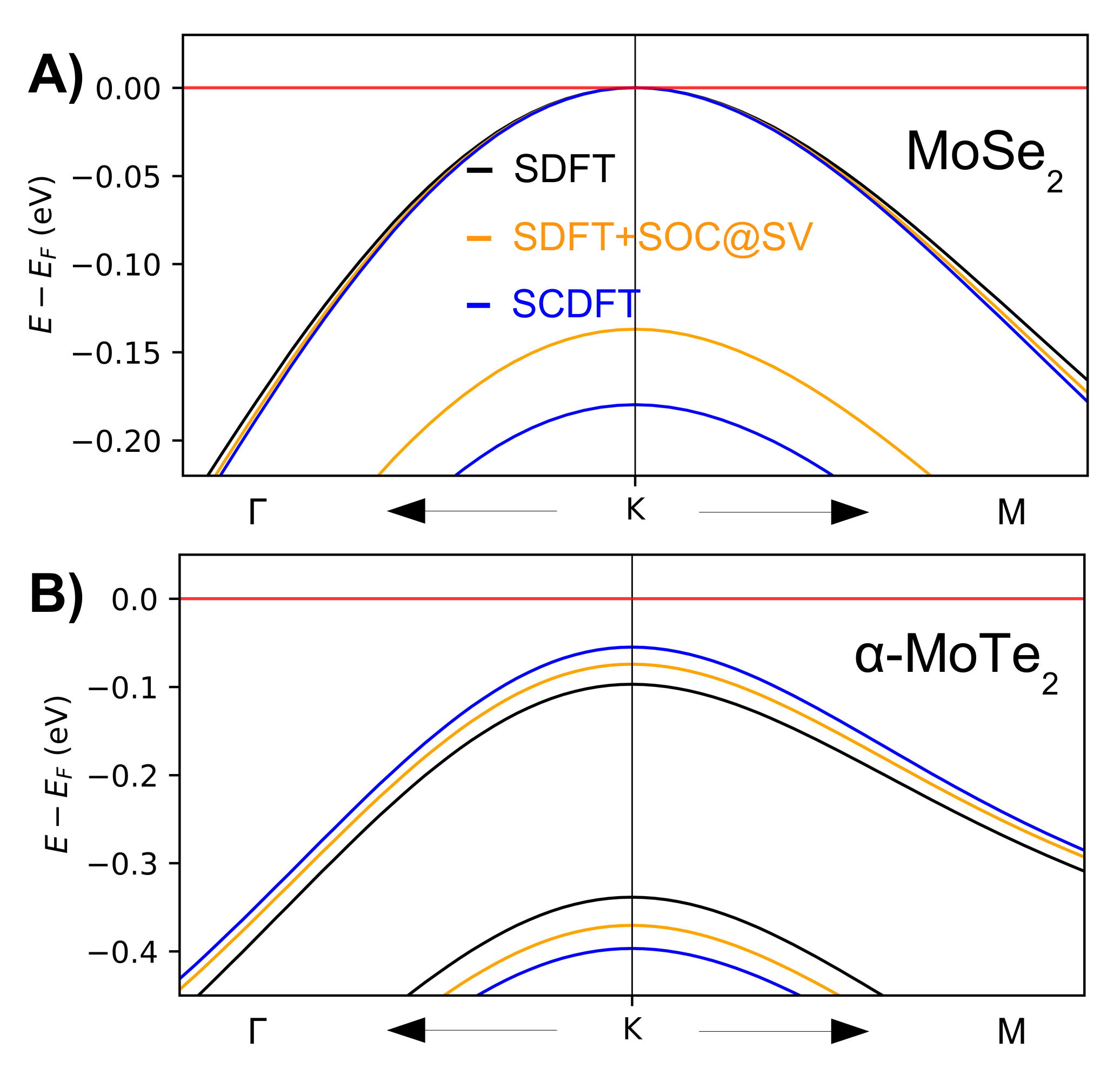}
\caption{SOC-induced/enhanced band splitting near the top of the valence band of the MoSe$_2$ single-layer obtained at the level of 
 a global hybrid (see text). Bands obtained from SDFT calculations (without SOC) are in black; from SDFT+SOC@SV  obtained by correcting the SDFT results by including SOC  in a second variational step  (in yellow); and those obtained from the present work, which accounts for SOC self-consistently, are in blue.  
The mixing parameter between Fock exchange and  PBE xc functional was  set  at 
$\alpha=0.15$.
Band structure images are produced with the CRYSTALpytools Python interface to \textsc{Crystal}.\cite{CRYSTALpytools}}
\label{fig:elstru}
\end{figure}

Let us discuss formal aspects of valence band splittings in the presence of SOC and different symmetry constraints. The  systems here considered preserve time-reversal symmetry (TRS):
\begin{equation}
\label{eq:TRS}
\varepsilon_k^\uparrow({\bf k}) = \varepsilon_k^\downarrow({\bf -k}) \; ,
\end{equation}
where $\varepsilon_k$ are the energy values of band $k$ at different points of the first Brillouin zone (FBZ). Let us recall that space inversion symmetry (SIS) results in the following constraint on the band structure:
\begin{equation}
\label{eq:SIS}
\varepsilon_k^\sigma({\bf k}) = \varepsilon_k^\sigma({\bf -k}) \; .
\end{equation}

 TRS and SIS together imply bands which are doubly degenerate in spin.
The inclusion of SOC makes the Hamiltonian spin-dependent, correspondingly spin-up and spin-down states feel a different potential and split, if allowed by symmetry. 

{\em The single-layer MoSe$_2$ system} preserves TRS but breaks SIS, which, in the presence of SOC, leads to possible spin-splittings of bands that would otherwise be doubly-degenerate at the SDFT level:
\begin{equation}
\label{eq:RasI}
\varepsilon_k^\uparrow({\bf k}) \neq \varepsilon_k^\downarrow({\bf k}) \; .
\end{equation}
In uniaxial (or low-dimensional) systems, such as 2D hexagonal MoSe$_2$, the spin-splittings are embodied by the Rashba Hamiltonian (Rashba-I effect).\cite{rashba1960properties} Figure \ref{fig:elstru} A) shows such spin-splitting  at the high-symmetry point K of the FBZ and along K-$\Gamma$ and K-M paths.  At the SDFT level (black line), the top valence band is doubly degenerate. The spin degeneracy is lifted by SOC according to Eq. (\ref{eq:RasI}); for instance, see the SCDFT description (blue lines).  

{\em The $\alpha$-MoTe$_2$ hexagonal crystal} is characterized by stacked MoTe$_2$ layers along the \textbf{c} crystallographic axis, separated by van-der-Waals gaps. As both TRS and SIS are preserved, the combination of Eqs. (\ref{eq:TRS}) and (\ref{eq:SIS}) leads to:
\begin{equation}
\label{eq:RasII}
\varepsilon_k^\uparrow({\bf k}) = \varepsilon_k^\downarrow({\bf k}) \; ,
\end{equation}
so that all bands are necessarily spin degenerate. 
Therefore, in the case of $\alpha$-MoTe$_2$, SOC enhanced band splittings are related to the dipole field of the locally asymmetric Mo crystallographic sites. This so-called Rashba-II effect results in the appearance of spatially localized ``hidden spin valleys'' associated with the band splittings.\cite{zhang2014hidden,oliva2020hidden}
Figure \ref{fig:elstru} B) shows such enhanced band splitting around K in the top of the valence band. At the SDFT level (black lines), the two top valence bands are spin doubly degenerate and are already split. With SOC, the bands are still doubly degenerate according to Eq. (\ref{eq:RasII}) but get further split (see the SCDFT blue lines).

\begin{figure}[t!]
\centering
\includegraphics[width=8.6cm]{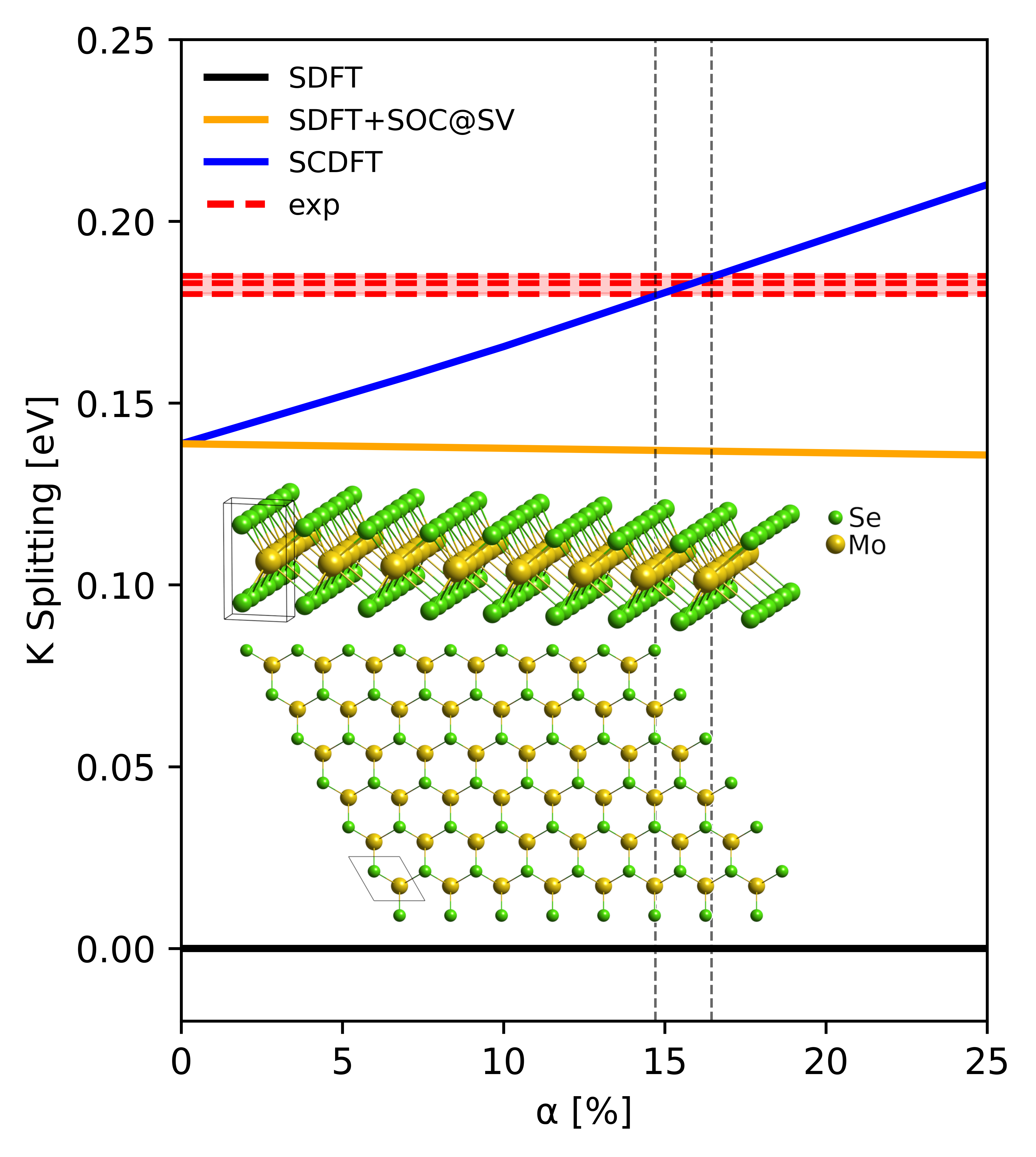}
\caption{Rashba-I type SOC-induced spin-splitting at the K point of the FBZ of 2D single-layer MoSe$_2$
obtained at the level of  a global hybrid (see text).
At the SDFT level (black line), SOC is not included. Yellow
and blue lines describe computed spin-splittings from different treatments of SOC as a function of $\alpha$ (i.e. fraction of Fock exchange): SDFT+SOC second-variational (yellow line)
and SCDFT (blue line), respectively. Experimental data (red lines) are taken from Refs. \onlinecite{zhang2014direct,shim2014large,ross2013electrical}. The atomic structure of the system is also shown.}
\label{fig:mosese}
\end{figure}

We perform calculations with the \textsc{Crystal23} package.\cite{Erba2023} We first report on  calculations employing a global-hybrid functional
as in Eq. \eqref{rDFA} {, making sure to allow for unrestricted  two-component single particle spinors,} with the PBE generalized-gradient approximation.\cite{perdew1996generalized} Computational details are reported in the supplementary material.\cite{ESI} (see also Refs. \onlinecite{desmarais2019spinI,doll2001analytical,doll2001implementation,doll2006analytical,civalleri2001hartree,metz2000small,peterson2003systematically,peterson2007energy,laun2018consistent,heyd2005energy,lebedev1976quadratures,lebedev1977spherical,towler1996density} therein). 
Below calculations are also performed and results reported for  a regular GGA (PBE) and
for the spin current-dependent MGGA (JLP) as defined in Eq.
\eqref{H@JLP}.


We start by discussing Rashba-I type SOC-induced spin-splitting in 2D single-layer MoSe$_2$. A graphical representation (side and top views) of the atomic structure of this system is given in Figure \ref{fig:mosese}. The splitting occurs at the K point of the FBZ of the system and has been measured by angle-resolved photo-electron spectroscopy (ARPES) experiments (0.180-0.185 eV).\cite{zhang2014direct,shim2014large,ross2013electrical} 

Figure \ref{fig:mosese} reports the computed spin-splittings as a function of $\alpha$ (i.e. fraction of Fock exchange). 
All geometries were fully optimized for each value of $\alpha$.

 The following is observed:
(i) The black line reports the SDFT results: {\em no} spin-splitting is observed, as expected; 
(ii) The yellow lines describes the results from  one-shot second-variational treatment of SOC (SDFT+SOC@SV). A value of 0.14 eV is obtained that significantly underestimates the experimental values, (almost) independently of $\alpha$.; 
iii) The blue line shows  the results from SCDFT calculations. The experimental band splittings are reproduced at values of $\alpha$ in the range 14-17\%. Correspondingly, the effect on the band splitting is found to amount to 22\% of the total SOC-induced splitting.

\begin{figure}[t!]
\centering
\includegraphics[width=8.6cm]{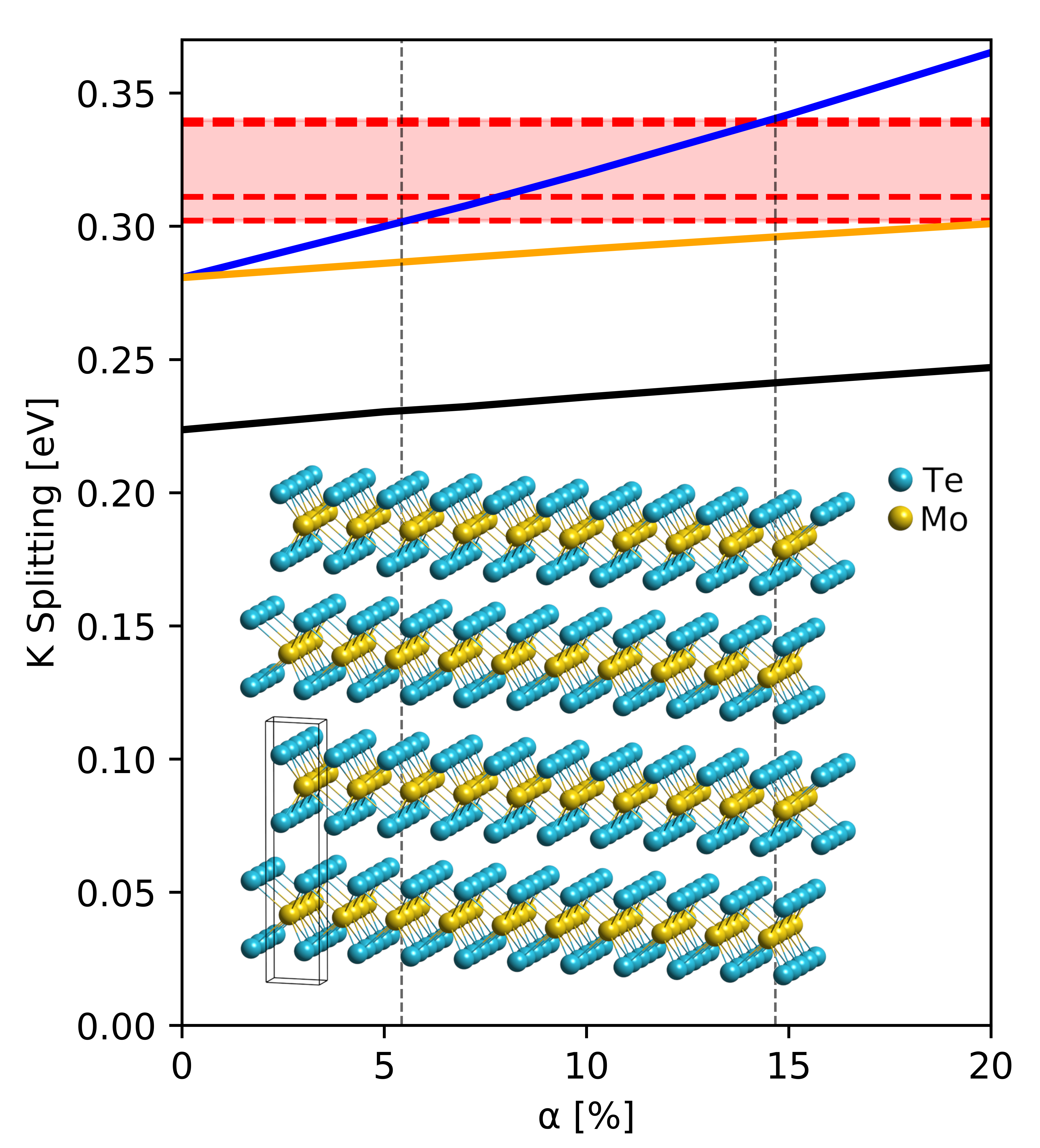}
\caption{As figure \ref{fig:mosese} but for Rashba-II type SOC-enhanced band-splitting at the K point of the FBZ of bulk $\alpha$-MoTe$_2$
obtained at the level of  a global hybrid (see text). Experimental data are taken from Refs. \onlinecite{beal1972transmission,oliva2020hidden,ruppert2014optical}. The atomic structure of the system is also shown.}
\label{fig:motete}
\end{figure}

\begin{table}[t!]
\caption{Single layer MoSe$_2$: SOC induced splittings (in eV), employing a pure GGA ($\alpha=0$) or hybrid and (J)MGGA functionals ($\alpha=0.15$). Geometries were optimized with the hybrid functional  at the SDFT level.}
\label{tab:mosese_split}
\begin{tabular}{ccccc}
      \hline
    \hline
         & GGA & MGGA & Hybrid & JMGGA \\
        \hline \hline\\
        SDFT    &  0 & 0 & 0  &  0 \\
        SDFT+SOC@SV & 0.14 & 0.14 &  0.14 &  0.14 \\
        SCDFT  & 0.14 & 0.14 & 0.18   & 0.18 \\
             Exp.   & & & 0.180 - 0.185 &  \\
    \hline
    \hline
    \end{tabular}
\end{table}

\begin{table}[t!]
\caption{Bulk $\alpha$-MoTe$_2$: SOC enhancement of band splittings (in eV) , employing a pure GGA ($\alpha=0$) or hybrid and (J)MGGA functionals ($\alpha=0.10$). Geometries were optimized with the hybrid functional at the SDFT level.}\label{tab:motete_split}
\begin{tabular}{ccccc}
        \hline
    \hline
        & GGA & MGGA & Hybrid & {JMGGA} \\
        \hline \hline\\
        SDFT    &  0.22 & 0.25 & 0.23  & 0.25 \\
        SDFT+SOC@SV & 0.22 & 0.31 &  0.29 &  0.31 \\
        SCDFT  & 0.22 & 0.31 & 0.32 & 0.33 \\
             Exp.   & & & 0.30 - 0.34 &  \\     
    \hline
    \hline
    \end{tabular}
\end{table}

\begin{table}[t!]
\caption{ SCDFT fundamental band gap $E_g$ (in eV). Experimental data are from Refs. \onlinecite{choi2017temperature,island2016precise,zelewski2017photoacoustic,ruppert,conan1979temperature}. 
 Computed values are reported for $\alpha=0.15$ for MoSe$_2$ and $\alpha=0.10$ for $\alpha$-MoTe$_2$. Geometries were optimized with the  hybrid functional at the SDFT level.}\label{tab:gaps}
\vspace{5pt}
\begin{tabular}{ccc}
    \hline
    \hline
              & {MoSe$_2$} & {MoTe$_2$} \\
    &&\\
        GGA    & 1.47 &  0.71  \\
        Hybrid   &  1.99 & 1.05   \\
        {JMGGA} &  1.65 & 0.71   \\
        &&\\
        Exp.   &  1.6-2.3 & 1.03   \\
    \hline
    \hline
    \end{tabular}
\end{table}


Next, we discuss Rashba-II type SOC-enhanced band-splitting in bulk $\alpha$-MoTe$_2$. A graphical representation of the atomic structure of this system is given in Figure \ref{fig:motete}. 
Here, two spin doubly-degenerate bands near the top of the valence are already split at the SDFT level (black lines) at the K point of the FBZ. The energy gap further widens upon inclusion of SOC, by an extent that depends on how spin-currents are treated.

The splitting has been measured by optical experiments (0.30-0.34 eV).\cite{beal1972transmission,oliva2020hidden,ruppert2014optical} 
Figure \ref{fig:motete} compares the experimental values with computed band-splittings from different treatments of SOC as a function of $\alpha$ (i.e. fraction of Fock exchange). We note that, for this system, experimental values are more significantly spread, which results in a more difficult quantitative assessment of the different theoretical approaches. However, the following is observed: i)   SDFT values visibly underestimate  experimental results; ii)   The SDFT+SOC@SV  
results are better than the SDFT results, as expected;
iii) The slope of the SDFT(+SOC@SV) results, however, is significantly different from the slope of the  SCDFT results. As a consequence of which,  agreement for the band splittings is obtained at an $\alpha$ which does {\em not} yield a band gap in agreement with experiments. Indeed, the SV calculation at a fraction $\alpha=0.2$ provides a splitting of 0.30 eV, but the band gap is much too large at 1.47 eV.; 
iv) The experimental band splitting is reproduced via SCDFT calculations at values of $\alpha$ in the range 0.06-0.15. It amounts to about 20\% of the total band-splitting at a fraction $\alpha=0.10$.

Tables \ref{tab:mosese_split} -- \ref{tab:gaps} summarize our results on splittings and fundamental band gaps for MoSe$_2$ and MoTe$_2$, employing, respectively, fractions $\alpha=0.15$ and $\alpha=0.10$ of Fock exchange. We reiterate that by employing one and the same hybrid form, 
we can reproduce the experimental band gaps and splittings with one and the same value of $\alpha$ (splittings of 0.18 and 0.32 eV, gaps of 1.99 and 1.05 eV, respectively, on MoSe$_2$ and MoTe$_2$).

We also stress that the  difference in the slopes between the yellow and blue lines in Fig. \ref{fig:mosese} and Fig. \ref{fig:motete} can unambiguously be attributed to the different treatment of the spin-currents.
Such a dependence  is implicitly encoded in the Fock exchange and it can be exploited by taking as an input  spinors derived under the action of SOC.
SDFT, however,  neglects SOC from the outset; so spin currents  vanish in the corresponding solutions.
SDFT+SOC@SV accounts for SOC but 
evaluates  Fock exchange at the level of  SDFT spinors; thus, after the second variation step, spin-currents do not vanish but are not used as a feedback in the calculation.
SCDFT, by construction,  evaluates Fock exchange under the action of SOC; thus spin currents can drive the convergence toward more accurate 
{\em self-consistent} results.

Finally, we pass to the GGA and the JMGGA cases. 
Not surprisingly, of course, neither the experimental gaps nor the splittings can be reproduced with a regular GGA functional.
Eq. \eqref{H@JLP} via Eqs. \eqref{JLP} and \eqref{AJLP} makes explicit the dependence of the exchange energy on spin-currents {\em and} brings forth the corresponding non-Abelian exchange potential. {\em JMGGA lowers the cost of the analogous global hybrid calculations --- by reducing Eq. \eqref{rDFA}  to \eqref{H@JLP} --- yet maintaining accurate band splittings (0.18 and 0.33 eV).} In doing so, however, the fundamental band gaps are decreased (i.e. from 1.99 to 1.65 eV on MoSe$_2$, from 1.05 to 0.71 eV on MoTe$_2$), which worsens the agreement against the experiment, when compared to the results of the full hybrid approximation. Nonetheless, the JMGGA gap of 1.65 eV on MoSe$_2$ is still an improvement over the pure GGA value of 1.47 eV. SDFT+SOC@SV calculations underestimate splittings (0.14 eV on MoSe$_2$, 0.22-0.31 eV on MoTe$_2$).

In conclusion, among the cases here considered, only the non-local Fock potential allows for a simultaneous agreement against the experiment on both fundamental band gaps and band splittings. If only band splittings are required, comparably good results can be obtained by replacing the non-local Fock operator by a computationally cheaper and formally simpler semi-local spin-current dependent approximation.

\subsection{Near-future road-map for GKS-SCDFT}

The results illustrated  above show that GKS-SCDFT is readily useful. Two questions can be posed, however: (i) Will it be possible to get rid of the empiricism involved in the determination of $\alpha$, the ``optimal'' fraction of exchange; or -- we may ask --- can the fraction be determined self-consistently in SCDFT without having to resort to 
other (computationally more demanding) methodologies? (ii) Will it be possible to derive more accurate functional approximations with an explicit dependence on the spin-current?
We foresee that the answer to both questions is likely to be positive. 

Question (i) may be resolved by upgrading a very recent development for 
 optimally-tuned range separated hybrids,\cite{Wing2021} which has been shown to work both for molecular and periodic materials.

Question (ii) may be answered by invoking the U(1)$\times$SU(2)-gauge invariant extension\cite{Pittalis2017} of more evolved meta-GGAs than the case reported here for illustrative purposes.
Work is in progress on the SCAN\cite{SCAN} and TASK\cite{TASK} energy functionals and the like.

\section{Outlooks and conclusions}
\label{Sec5}



We have put forward a generalization  of the Kohn-Sham formalism (GKS), which admits the use of non-local effective potentials firmly rooted in SCDFT. This formulation is the analogous of the popular GKS formulation of (Spin-)DFT.\cite{Seidl1996} 
Here, we have spelled out and analyzed the novel and subtle aspects that are uniquely brought forth by the SCDFT framework.  
We have  demonstrated via applications that GKS-SCDFT readily allows us to obtain results beyond the state-of-the-art in electronic structure calculations for spin-orbit coupled materials. By considering time-reversal symmetric spin-orbit coupled states, we have demonstrated that the  dependence of the energy functional on spin currents is important even when it is only implicit, as in the prominent case of Fock exchange. Global hybrid  approximations can yield  significantly more accurate results when used in GKS-SCDFT  calculations rather than in perturbative SDFT+SOC calculations.

In particular, we have applied GKS-SCDFT to the evaluation
of band gaps {\em and} SOC-induced band splittings in materials of great interest in spintronics and valleytronics, with Rashba-I, Rashba-II effects. 
At the level of the global hybrid approximations, we  have shown that by applying the self-consistent SCDFT treatment of spin-orbit interactions one can find an optimal fraction of  Fock exchange which works well for both the fundamental band gaps {\em and} the SOC-induced or -enhanced band splittings.
We have shown that the widely used method for  refining Spin-DFT  results via a second-variational treatment of SOC can
fail to reproduce the experimental results -- superior agreement can be achieved by switching from SDFT to full-fledged SCDFT calculations. 

Efforts, in the near future,  will be devoted to reduce empiricism  in finding the ``optimal'' fraction of Fock exchange. We believe that the optimally-tuned range separated hybrids offer, presently, a valid and very promising option.~\cite{Wing2021} 
Furthermore, the illustrative case reported here of a simple spin-current dependent meta-GGA suggests that it is appealing  to develop the
 extension of more recent and more evolved meta-GGA forms to SCDFT as well.

Most importantly, already at this stage of the development, the GKS approach of Spin-Current DFT can  offer  significant improvements in the calculation of the electronic structure of challenging spin-orbit coupled materials.\\

\begin{acknowledgments}
This research has received funding from the Project CH4.0 under the MUR program ``Dipartimenti di Eccellenza 2023-2027'' (CUP: D13C22003520001). GV was supported by the Ministry of Education, Singapore, under its Research Centre of Excellence award to the Institute for Functional Intelligent Materials (I-FIM, project No. EDUNC-33-18-279-V12). We are grateful to Stephen Dale for a reading of the manuscript.
\end{acknowledgments}

\appendix

\bibliography{paper_resubmit_arxiv}

\begin{thebibliography}{101}
\expandafter\ifx\csname natexlab\endcsname\relax\def\natexlab#1{#1}\fi
\expandafter\ifx\csname bibnamefont\endcsname\relax
  \def\bibnamefont#1{#1}\fi
\expandafter\ifx\csname bibfnamefont\endcsname\relax
  \def\bibfnamefont#1{#1}\fi
\expandafter\ifx\csname citenamefont\endcsname\relax
  \def\citenamefont#1{#1}\fi
\expandafter\ifx\csname url\endcsname\relax
  \def\url#1{\texttt{#1}}\fi
\expandafter\ifx\csname urlprefix\endcsname\relax\def\urlprefix{URL }\fi
\providecommand{\bibinfo}[2]{#2}
\providecommand{\eprint}[2][]{\url{#2}}

\bibitem[{\citenamefont{Greiner}(2012)}]{greiner2012relativistic}
\bibinfo{author}{\bibfnamefont{W.}~\bibnamefont{Greiner}},
  \emph{\bibinfo{title}{Relativistic Quantum Mechanics: Wave Equations}}
  (\bibinfo{publisher}{Springer Berlin Heidelberg}, \bibinfo{year}{2012}), ISBN
  \bibinfo{isbn}{9783642880827}.

\bibitem[{\citenamefont{Yu and Cardona}(2010)}]{yu2010fundamentals}
\bibinfo{author}{\bibfnamefont{P.}~\bibnamefont{Yu}} \bibnamefont{and}
  \bibinfo{author}{\bibfnamefont{M.}~\bibnamefont{Cardona}},
  \emph{\bibinfo{title}{Fundamentals of Semiconductors: Physics and Materials
  Properties}}, Graduate Texts in Physics (\bibinfo{publisher}{Springer Berlin
  Heidelberg}, \bibinfo{year}{2010}), ISBN \bibinfo{isbn}{9783642007101}.

\bibitem[{\citenamefont{Bernevig and Hughes}(2013)}]{bernevig2013topological}
\bibinfo{author}{\bibfnamefont{B.}~\bibnamefont{Bernevig}} \bibnamefont{and}
  \bibinfo{author}{\bibfnamefont{T.}~\bibnamefont{Hughes}},
  \emph{\bibinfo{title}{Topological Insulators and Topological
  Superconductors}} (\bibinfo{publisher}{Princeton University Press},
  \bibinfo{year}{2013}), ISBN \bibinfo{isbn}{9780691151755}.

\bibitem[{\citenamefont{Vanderbilt}(2018)}]{vanderbilt2018berry}
\bibinfo{author}{\bibfnamefont{D.}~\bibnamefont{Vanderbilt}},
  \emph{\bibinfo{title}{Berry Phases in Electronic Structure Theory: Electric
  Polarization, Orbital Magnetization and Topological Insulators}}
  (\bibinfo{publisher}{Cambridge University Press}, \bibinfo{year}{2018}), ISBN
  \bibinfo{isbn}{9781107157651}.

\bibitem[{\citenamefont{\ifmmode \check{Z}\else
  \v{Z}\fi{}uti\ifmmode~\acute{c}\else \'{c}\fi{}
  et~al.}(2004)\citenamefont{\ifmmode \check{Z}\else
  \v{Z}\fi{}uti\ifmmode~\acute{c}\else \'{c}\fi{}, Fabian, and
  Das~Sarma}}]{RevModPhys.76.323}
\bibinfo{author}{\bibfnamefont{I.}~\bibnamefont{\ifmmode \check{Z}\else
  \v{Z}\fi{}uti\ifmmode~\acute{c}\else \'{c}\fi{}}},
  \bibinfo{author}{\bibfnamefont{J.}~\bibnamefont{Fabian}}, \bibnamefont{and}
  \bibinfo{author}{\bibfnamefont{S.}~\bibnamefont{Das~Sarma}},
  \bibinfo{journal}{Rev. Mod. Phys.} \textbf{\bibinfo{volume}{76}},
  \bibinfo{pages}{323} (\bibinfo{year}{2004}).

\bibitem[{\citenamefont{Zhang and Vignale}(2016)}]{ZG16}
\bibinfo{author}{\bibfnamefont{S.~S.-L.} \bibnamefont{Zhang}} \bibnamefont{and}
  \bibinfo{author}{\bibfnamefont{G.}~\bibnamefont{Vignale}},
  \bibinfo{journal}{Phys. Rev. B} \textbf{\bibinfo{volume}{94}},
  \bibinfo{pages}{140411} (\bibinfo{year}{2016}).

\bibitem[{\citenamefont{Vignale and Rasolt}(1988)}]{VignaleRasolt:88}
\bibinfo{author}{\bibfnamefont{G.}~\bibnamefont{Vignale}} \bibnamefont{and}
  \bibinfo{author}{\bibfnamefont{M.}~\bibnamefont{Rasolt}},
  \bibinfo{journal}{Phys. Rev. B} \textbf{\bibinfo{volume}{37}},
  \bibinfo{pages}{10685} (\bibinfo{year}{1988}).

\bibitem[{\citenamefont{Bencheikh}(2003)}]{Bencheikh:03}
\bibinfo{author}{\bibfnamefont{K.}~\bibnamefont{Bencheikh}},
  \bibinfo{journal}{J. Phys. A} \textbf{\bibinfo{volume}{36}},
  \bibinfo{pages}{11929} (\bibinfo{year}{2003}).

\bibitem[{\citenamefont{Rohra and G\"orling}(2006)}]{Goerling2006}
\bibinfo{author}{\bibfnamefont{S.}~\bibnamefont{Rohra}} \bibnamefont{and}
  \bibinfo{author}{\bibfnamefont{A.}~\bibnamefont{G\"orling}},
  \bibinfo{journal}{Phys. Rev. Lett.} \textbf{\bibinfo{volume}{97}},
  \bibinfo{pages}{013005} (\bibinfo{year}{2006}).

\bibitem[{\citenamefont{Heaton-Burgess
  et~al.}(2007)\citenamefont{Heaton-Burgess, Ayers, and
  Yang}}]{HeatonBurgess2007}
\bibinfo{author}{\bibfnamefont{T.}~\bibnamefont{Heaton-Burgess}},
  \bibinfo{author}{\bibfnamefont{P.}~\bibnamefont{Ayers}}, \bibnamefont{and}
  \bibinfo{author}{\bibfnamefont{W.}~\bibnamefont{Yang}},
  \bibinfo{journal}{Phys. Rev. Lett.} \textbf{\bibinfo{volume}{98}},
  \bibinfo{pages}{036403} (\bibinfo{year}{2007}).

\bibitem[{\citenamefont{Abedinpour et~al.}(2010)\citenamefont{Abedinpour,
  Vignale, and Tokatly}}]{AbedinpourTokatly:10}
\bibinfo{author}{\bibfnamefont{S.~H.} \bibnamefont{Abedinpour}},
  \bibinfo{author}{\bibfnamefont{G.}~\bibnamefont{Vignale}}, \bibnamefont{and}
  \bibinfo{author}{\bibfnamefont{I.~V.} \bibnamefont{Tokatly}},
  \bibinfo{journal}{Phys. Rev. B} \textbf{\bibinfo{volume}{81}},
  \bibinfo{pages}{125123} (\bibinfo{year}{2010}).

\bibitem[{\citenamefont{Kurth and Pittalis}(2006)}]{Kurth2006}
\bibinfo{author}{\bibfnamefont{S.}~\bibnamefont{Kurth}} \bibnamefont{and}
  \bibinfo{author}{\bibfnamefont{S.}~\bibnamefont{Pittalis}},
  \emph{\bibinfo{title}{{T}he {O}ptimized {E}ffective {P}otential {M}ethod and
  {LDA} + {U}}} (\bibinfo{publisher}{John von Neumann Institute for Computing},
  \bibinfo{address}{Jülich}, \bibinfo{year}{2006}), vol.~\bibinfo{volume}{31}
  of \emph{\bibinfo{series}{NIC series}}, pp. \bibinfo{pages}{299--334}.

\bibitem[{\citenamefont{Pittalis et~al.}(2006)\citenamefont{Pittalis, Kurth,
  Helbig, and Gross}}]{Pittalis2006}
\bibinfo{author}{\bibfnamefont{S.}~\bibnamefont{Pittalis}},
  \bibinfo{author}{\bibfnamefont{S.}~\bibnamefont{Kurth}},
  \bibinfo{author}{\bibfnamefont{N.}~\bibnamefont{Helbig}}, \bibnamefont{and}
  \bibinfo{author}{\bibfnamefont{E.~K.~U.} \bibnamefont{Gross}},
  \bibinfo{journal}{Phys. Rev. A} \textbf{\bibinfo{volume}{74}},
  \bibinfo{pages}{062511} (\bibinfo{year}{2006}).

\bibitem[{\citenamefont{Sharma et~al.}(2007{\natexlab{a}})\citenamefont{Sharma,
  Dewhurst, Ambrosch-Draxl, Kurth, Helbig, Pittalis, Shallcross, Nordstr\"om,
  and Gross}}]{Sharma2007a}
\bibinfo{author}{\bibfnamefont{S.}~\bibnamefont{Sharma}},
  \bibinfo{author}{\bibfnamefont{J.~K.} \bibnamefont{Dewhurst}},
  \bibinfo{author}{\bibfnamefont{C.}~\bibnamefont{Ambrosch-Draxl}},
  \bibinfo{author}{\bibfnamefont{S.}~\bibnamefont{Kurth}},
  \bibinfo{author}{\bibfnamefont{N.}~\bibnamefont{Helbig}},
  \bibinfo{author}{\bibfnamefont{S.}~\bibnamefont{Pittalis}},
  \bibinfo{author}{\bibfnamefont{S.}~\bibnamefont{Shallcross}},
  \bibinfo{author}{\bibfnamefont{L.}~\bibnamefont{Nordstr\"om}},
  \bibnamefont{and} \bibinfo{author}{\bibfnamefont{E.~K.~U.}
  \bibnamefont{Gross}}, \bibinfo{journal}{Phys. Rev. Lett.}
  \textbf{\bibinfo{volume}{98}}, \bibinfo{pages}{196405}
  (\bibinfo{year}{2007}{\natexlab{a}}).

\bibitem[{\citenamefont{Sharma et~al.}(2007{\natexlab{b}})\citenamefont{Sharma,
  Pittalis, Kurth, Shallcross, Dewhurst, and Gross}}]{Sharma2007b}
\bibinfo{author}{\bibfnamefont{S.}~\bibnamefont{Sharma}},
  \bibinfo{author}{\bibfnamefont{S.}~\bibnamefont{Pittalis}},
  \bibinfo{author}{\bibfnamefont{S.}~\bibnamefont{Kurth}},
  \bibinfo{author}{\bibfnamefont{S.}~\bibnamefont{Shallcross}},
  \bibinfo{author}{\bibfnamefont{J.~K.} \bibnamefont{Dewhurst}},
  \bibnamefont{and} \bibinfo{author}{\bibfnamefont{E.~K.~U.}
  \bibnamefont{Gross}}, \bibinfo{journal}{Phys. Rev. B}
  \textbf{\bibinfo{volume}{76}}, \bibinfo{pages}{100401}
  (\bibinfo{year}{2007}{\natexlab{b}}).

\bibitem[{\citenamefont{Trushin and G\"orling}(2018)}]{Goerling2018}
\bibinfo{author}{\bibfnamefont{E.}~\bibnamefont{Trushin}} \bibnamefont{and}
  \bibinfo{author}{\bibfnamefont{A.}~\bibnamefont{G\"orling}},
  \bibinfo{journal}{Phys. Rev. B} \textbf{\bibinfo{volume}{98}},
  \bibinfo{pages}{205137} (\bibinfo{year}{2018}).

\bibitem[{\citenamefont{Ullrich}(2018)}]{Ullrich2018}
\bibinfo{author}{\bibfnamefont{C.~A.} \bibnamefont{Ullrich}},
  \bibinfo{journal}{Phys. Rev. B} \textbf{\bibinfo{volume}{98}},
  \bibinfo{pages}{035140} (\bibinfo{year}{2018}).

\bibitem[{\citenamefont{Pluhar and Ullrich}(2019)}]{Pluhar2019}
\bibinfo{author}{\bibfnamefont{E.~A.} \bibnamefont{Pluhar}} \bibnamefont{and}
  \bibinfo{author}{\bibfnamefont{C.~A.} \bibnamefont{Ullrich}},
  \bibinfo{journal}{Phys. Rev. B} \textbf{\bibinfo{volume}{100}},
  \bibinfo{pages}{125135} (\bibinfo{year}{2019}).

\bibitem[{\citenamefont{Pittalis et~al.}(2017)\citenamefont{Pittalis, Vignale,
  and Eich}}]{Pittalis2017}
\bibinfo{author}{\bibfnamefont{S.}~\bibnamefont{Pittalis}},
  \bibinfo{author}{\bibfnamefont{G.}~\bibnamefont{Vignale}}, \bibnamefont{and}
  \bibinfo{author}{\bibfnamefont{F.~G.} \bibnamefont{Eich}},
  \bibinfo{journal}{Phys. Rev. B} \textbf{\bibinfo{volume}{96}},
  \bibinfo{pages}{035141} (\bibinfo{year}{2017}).

\bibitem[{\citenamefont{WU and YANG}(2003)}]{QIN2003}
\bibinfo{author}{\bibfnamefont{Q.}~\bibnamefont{WU}} \bibnamefont{and}
  \bibinfo{author}{\bibfnamefont{W.}~\bibnamefont{YANG}}, \bibinfo{journal}{J.
  Theor. Comput. Chem.} \textbf{\bibinfo{volume}{02}}, \bibinfo{pages}{627}
  (\bibinfo{year}{2003}).

\bibitem[{\citenamefont{Gidopoulos and Lathiotakis}(2012)}]{Gidopoulos2012}
\bibinfo{author}{\bibfnamefont{N.~I.} \bibnamefont{Gidopoulos}}
  \bibnamefont{and} \bibinfo{author}{\bibfnamefont{N.~N.}
  \bibnamefont{Lathiotakis}}, \bibinfo{journal}{Phys. Rev. A}
  \textbf{\bibinfo{volume}{85}}, \bibinfo{pages}{052508}
  (\bibinfo{year}{2012}).

\bibitem[{\citenamefont{Eich and Hellgren}(2014)}]{Eich2014}
\bibinfo{author}{\bibfnamefont{F.~G.} \bibnamefont{Eich}} \bibnamefont{and}
  \bibinfo{author}{\bibfnamefont{M.}~\bibnamefont{Hellgren}},
  \bibinfo{journal}{J. Chem. Phys.} \textbf{\bibinfo{volume}{141}}
  (\bibinfo{year}{2014}).

\bibitem[{\citenamefont{Gidopoulos and Lathiotakis}(2013)}]{Gidopoulos2013}
\bibinfo{author}{\bibfnamefont{N.~I.} \bibnamefont{Gidopoulos}}
  \bibnamefont{and} \bibinfo{author}{\bibfnamefont{N.~N.}
  \bibnamefont{Lathiotakis}}, \bibinfo{journal}{Phys. Rev. A}
  \textbf{\bibinfo{volume}{88}}, \bibinfo{pages}{046502}
  (\bibinfo{year}{2013}).

\bibitem[{\citenamefont{Oueis et~al.}(2023)\citenamefont{Oueis, Sizov, and
  Staroverov}}]{Staroverov2023}
\bibinfo{author}{\bibfnamefont{Y.}~\bibnamefont{Oueis}},
  \bibinfo{author}{\bibfnamefont{G.~N.} \bibnamefont{Sizov}}, \bibnamefont{and}
  \bibinfo{author}{\bibfnamefont{V.~N.} \bibnamefont{Staroverov}},
  \bibinfo{journal}{J. Phys. Chem. A} \textbf{\bibinfo{volume}{127}},
  \bibinfo{pages}{2664} (\bibinfo{year}{2023}).

\bibitem[{\citenamefont{Becke}(1993)}]{becke1993new}
\bibinfo{author}{\bibfnamefont{A.~D.} \bibnamefont{Becke}},
  \bibinfo{journal}{J. Chem. Phys.} \textbf{\bibinfo{volume}{98}},
  \bibinfo{pages}{1372} (\bibinfo{year}{1993}).

\bibitem[{\citenamefont{Seidl et~al.}(1996)\citenamefont{Seidl, G\"orling,
  Vogl, Majewski, and Levy}}]{Seidl1996}
\bibinfo{author}{\bibfnamefont{A.}~\bibnamefont{Seidl}},
  \bibinfo{author}{\bibfnamefont{A.}~\bibnamefont{G\"orling}},
  \bibinfo{author}{\bibfnamefont{P.}~\bibnamefont{Vogl}},
  \bibinfo{author}{\bibfnamefont{J.~A.} \bibnamefont{Majewski}},
  \bibnamefont{and} \bibinfo{author}{\bibfnamefont{M.}~\bibnamefont{Levy}},
  \bibinfo{journal}{Phys. Rev. B} \textbf{\bibinfo{volume}{53}},
  \bibinfo{pages}{3764} (\bibinfo{year}{1996}).

\bibitem[{\citenamefont{Perdew et~al.}(2017)\citenamefont{Perdew, Yang, Burke,
  Yang, Gross, Scheffler, Scuseria, Henderson, Zhang, Ruzsinszky
  et~al.}}]{perdew2017understanding}
\bibinfo{author}{\bibfnamefont{J.~P.} \bibnamefont{Perdew}},
  \bibinfo{author}{\bibfnamefont{W.}~\bibnamefont{Yang}},
  \bibinfo{author}{\bibfnamefont{K.}~\bibnamefont{Burke}},
  \bibinfo{author}{\bibfnamefont{Z.}~\bibnamefont{Yang}},
  \bibinfo{author}{\bibfnamefont{E.~K.} \bibnamefont{Gross}},
  \bibinfo{author}{\bibfnamefont{M.}~\bibnamefont{Scheffler}},
  \bibinfo{author}{\bibfnamefont{G.~E.} \bibnamefont{Scuseria}},
  \bibinfo{author}{\bibfnamefont{T.~M.} \bibnamefont{Henderson}},
  \bibinfo{author}{\bibfnamefont{I.~Y.} \bibnamefont{Zhang}},
  \bibinfo{author}{\bibfnamefont{A.}~\bibnamefont{Ruzsinszky}},
  \bibnamefont{et~al.}, \bibinfo{journal}{PNAS} \textbf{\bibinfo{volume}{114}},
  \bibinfo{pages}{2801} (\bibinfo{year}{2017}).

\bibitem[{\citenamefont{Neumann et~al.}(1996)\citenamefont{Neumann, Nobes, and
  Handy}}]{Neumann1996}
\bibinfo{author}{\bibfnamefont{R.}~\bibnamefont{Neumann}},
  \bibinfo{author}{\bibfnamefont{R.~H.} \bibnamefont{Nobes}}, \bibnamefont{and}
  \bibinfo{author}{\bibfnamefont{N.~C.} \bibnamefont{Handy}},
  \bibinfo{journal}{Mol. Phys.} \textbf{\bibinfo{volume}{87}},
  \bibinfo{pages}{1} (\bibinfo{year}{1996}).

\bibitem[{\citenamefont{Lehtola et~al.}(2020)\citenamefont{Lehtola, Blockhuys,
  and Van~Alsenoy}}]{Lehtola2020}
\bibinfo{author}{\bibfnamefont{S.}~\bibnamefont{Lehtola}},
  \bibinfo{author}{\bibfnamefont{F.}~\bibnamefont{Blockhuys}},
  \bibnamefont{and}
  \bibinfo{author}{\bibfnamefont{C.}~\bibnamefont{Van~Alsenoy}},
  \bibinfo{journal}{Mol.} \textbf{\bibinfo{volume}{25}} (\bibinfo{year}{2020}).

\bibitem[{\citenamefont{Wing et~al.}(2021)\citenamefont{Wing, Ohad, Haber,
  Filip, Gant, Neaton, and Kronik}}]{Wing2021}
\bibinfo{author}{\bibfnamefont{D.}~\bibnamefont{Wing}},
  \bibinfo{author}{\bibfnamefont{G.}~\bibnamefont{Ohad}},
  \bibinfo{author}{\bibfnamefont{J.~B.} \bibnamefont{Haber}},
  \bibinfo{author}{\bibfnamefont{M.~R.} \bibnamefont{Filip}},
  \bibinfo{author}{\bibfnamefont{S.~E.} \bibnamefont{Gant}},
  \bibinfo{author}{\bibfnamefont{J.~B.} \bibnamefont{Neaton}},
  \bibnamefont{and} \bibinfo{author}{\bibfnamefont{L.}~\bibnamefont{Kronik}},
  \bibinfo{journal}{Proceedings of the National Academy of Sciences}
  \textbf{\bibinfo{volume}{118}}, \bibinfo{pages}{e2104556118}
  (\bibinfo{year}{2021}).

\bibitem[{\citenamefont{Desmarais
  et~al.}(2020{\natexlab{a}})\citenamefont{Desmarais, Flament, and
  Erba}}]{Desmarais2020a}
\bibinfo{author}{\bibfnamefont{J.~K.} \bibnamefont{Desmarais}},
  \bibinfo{author}{\bibfnamefont{J.-P.} \bibnamefont{Flament}},
  \bibnamefont{and} \bibinfo{author}{\bibfnamefont{A.}~\bibnamefont{Erba}},
  \bibinfo{journal}{Phys. Rev. B} \textbf{\bibinfo{volume}{101}},
  \bibinfo{pages}{235142} (\bibinfo{year}{2020}{\natexlab{a}}).

\bibitem[{\citenamefont{Desmarais
  et~al.}(2020{\natexlab{b}})\citenamefont{Desmarais, Flament, and
  Erba}}]{Desmarais2020b}
\bibinfo{author}{\bibfnamefont{J.~K.} \bibnamefont{Desmarais}},
  \bibinfo{author}{\bibfnamefont{J.-P.} \bibnamefont{Flament}},
  \bibnamefont{and} \bibinfo{author}{\bibfnamefont{A.}~\bibnamefont{Erba}},
  \bibinfo{journal}{Phys. Rev. B} \textbf{\bibinfo{volume}{102}},
  \bibinfo{pages}{235118} (\bibinfo{year}{2020}{\natexlab{b}}).

\bibitem[{\citenamefont{Desmarais
  et~al.}(2021{\natexlab{a}})\citenamefont{Desmarais, Komorovsky, Flament, and
  Erba}}]{desmarais2021spin2}
\bibinfo{author}{\bibfnamefont{J.~K.} \bibnamefont{Desmarais}},
  \bibinfo{author}{\bibfnamefont{S.}~\bibnamefont{Komorovsky}},
  \bibinfo{author}{\bibfnamefont{J.-P.} \bibnamefont{Flament}},
  \bibnamefont{and} \bibinfo{author}{\bibfnamefont{A.}~\bibnamefont{Erba}},
  \bibinfo{journal}{J. Chem. Phys.} \textbf{\bibinfo{volume}{154}},
  \bibinfo{pages}{204110} (\bibinfo{year}{2021}{\natexlab{a}}).

\bibitem[{\citenamefont{Desmarais et~al.}(2019)\citenamefont{Desmarais,
  Flament, and Erba}}]{desmarais2019spinI}
\bibinfo{author}{\bibfnamefont{J.~K.} \bibnamefont{Desmarais}},
  \bibinfo{author}{\bibfnamefont{J.-P.} \bibnamefont{Flament}},
  \bibnamefont{and} \bibinfo{author}{\bibfnamefont{A.}~\bibnamefont{Erba}},
  \bibinfo{journal}{J. Chem. Phys.} \textbf{\bibinfo{volume}{151}},
  \bibinfo{pages}{074107} (\bibinfo{year}{2019}).

\bibitem[{\citenamefont{Comaskey et~al.}(2022)\citenamefont{Comaskey, Bodo,
  Erba, Mendoza-Cortes, and Desmarais}}]{Comaskey2022}
\bibinfo{author}{\bibfnamefont{W.~P.} \bibnamefont{Comaskey}},
  \bibinfo{author}{\bibfnamefont{F.}~\bibnamefont{Bodo}},
  \bibinfo{author}{\bibfnamefont{A.}~\bibnamefont{Erba}},
  \bibinfo{author}{\bibfnamefont{J.~L.} \bibnamefont{Mendoza-Cortes}},
  \bibnamefont{and} \bibinfo{author}{\bibfnamefont{J.~K.}
  \bibnamefont{Desmarais}}, \bibinfo{journal}{Phys. Rev. B}
  \textbf{\bibinfo{volume}{106}}, \bibinfo{pages}{L201109}
  (\bibinfo{year}{2022}).

\bibitem[{\citenamefont{Bodo et~al.}(2022)\citenamefont{Bodo, Desmarais, and
  Erba}}]{Bodo2022}
\bibinfo{author}{\bibfnamefont{F.}~\bibnamefont{Bodo}},
  \bibinfo{author}{\bibfnamefont{J.~K.} \bibnamefont{Desmarais}},
  \bibnamefont{and} \bibinfo{author}{\bibfnamefont{A.}~\bibnamefont{Erba}},
  \bibinfo{journal}{Phys. Rev. B} \textbf{\bibinfo{volume}{105}},
  \bibinfo{pages}{125108} (\bibinfo{year}{2022}).

\bibitem[{\citenamefont{Fr\"ohlich and Studer}(1993)}]{FroehlichStuder:93}
\bibinfo{author}{\bibfnamefont{J.}~\bibnamefont{Fr\"ohlich}} \bibnamefont{and}
  \bibinfo{author}{\bibfnamefont{U.~M.} \bibnamefont{Studer}},
  \bibinfo{journal}{Rev. Mod. Phys.} \textbf{\bibinfo{volume}{65}},
  \bibinfo{pages}{733} (\bibinfo{year}{1993}).

\bibitem[{\citenamefont{Levy}(1982)}]{Levy:82}
\bibinfo{author}{\bibfnamefont{M.}~\bibnamefont{Levy}}, \bibinfo{journal}{Phys.
  Rev. A} \textbf{\bibinfo{volume}{26}}, \bibinfo{pages}{1200}
  (\bibinfo{year}{1982}).

\bibitem[{\citenamefont{Lieb}(1983)}]{Lieb:83}
\bibinfo{author}{\bibfnamefont{E.~H.} \bibnamefont{Lieb}},
  \bibinfo{journal}{Int. J. Quantum Chem.} \textbf{\bibinfo{volume}{24}},
  \bibinfo{pages}{243} (\bibinfo{year}{1983}).

\bibitem[{not()}]{note6}
\bibinfo{note}{We note that the mathematical conditions for
  $N$-representability and $v$-representability in SCDFT are presently unknown.
  Non-trivial notable differences with the case of standard DFT have recently
  been pointed out in SDFT, which make an ensemble formulation necessary even
  for $N$-representability~\cite{Gontier13}. Similar or bigger differences are
  expected in SCDFT. Thus, the domain of the $F$ functional in Eq.~(\ref{F}) is
  not known, and we have informally assumed that it is sufficiently dense to
  approximate most density sets that may occur in physical systems. Likewise,
  when introducing the Kohn-Sham system, we will informally assume
  non-interacting $v$-representability based on single Slater determinants. It
  is natural to expect that, the $N$-representability and $v$-representability
  issues which are prone to arise in SCDFT will be ameliorated (if not
  resolved) within an ensemble formulation, in which the admissible
  non-interacting pure states would also comprise linear combination of
  determinants.}

\bibitem[{\citenamefont{Garrick et~al.}(2022)\citenamefont{Garrick, Gould, and
  Kronik}}]{Garrick2022}
\bibinfo{author}{\bibfnamefont{R.}~\bibnamefont{Garrick}},
  \bibinfo{author}{\bibfnamefont{T.}~\bibnamefont{Gould}}, \bibnamefont{and}
  \bibinfo{author}{\bibfnamefont{L.}~\bibnamefont{Kronik}},
  \bibinfo{journal}{Adv. Theory Simul.} \textbf{\bibinfo{volume}{10}}
  (\bibinfo{year}{2022}).

\bibitem[{\citenamefont{Harris}(1984)}]{harris1984adiabatic}
\bibinfo{author}{\bibfnamefont{J.}~\bibnamefont{Harris}},
  \bibinfo{journal}{Phys. Rev. A} \textbf{\bibinfo{volume}{29}},
  \bibinfo{pages}{1648} (\bibinfo{year}{1984}).

\bibitem[{\citenamefont{Perdew et~al.}(1996{\natexlab{a}})\citenamefont{Perdew,
  Ernzerhof, and Burke}}]{perdew1996rationale}
\bibinfo{author}{\bibfnamefont{J.~P.} \bibnamefont{Perdew}},
  \bibinfo{author}{\bibfnamefont{M.}~\bibnamefont{Ernzerhof}},
  \bibnamefont{and} \bibinfo{author}{\bibfnamefont{K.}~\bibnamefont{Burke}},
  \bibinfo{journal}{J. Chem. Phys.} \textbf{\bibinfo{volume}{105}},
  \bibinfo{pages}{9982} (\bibinfo{year}{1996}{\natexlab{a}}).

\bibitem[{\citenamefont{Koch and Holthausen}(2nd edition,
  2001)}]{KochHolthausen2001}
\bibinfo{author}{\bibfnamefont{W.}~\bibnamefont{Koch}} \bibnamefont{and}
  \bibinfo{author}{\bibfnamefont{M.~C.} \bibnamefont{Holthausen}},
  \emph{\bibinfo{title}{{A} {C}hemist's {G}uide to {D}ensity {F}unctional
  {T}heory}} (\bibinfo{publisher}{Wiley - {VCH}}, \bibinfo{address}{Weinheim -
  New York}, \bibinfo{year}{2nd edition, 2001}).

\bibitem[{\citenamefont{Marques et~al.}(2011)\citenamefont{Marques, Vidal,
  Oliveira, Reining, and Botti}}]{Marques2011}
\bibinfo{author}{\bibfnamefont{M.~A.~L.} \bibnamefont{Marques}},
  \bibinfo{author}{\bibfnamefont{J.}~\bibnamefont{Vidal}},
  \bibinfo{author}{\bibfnamefont{M.~J.~T.} \bibnamefont{Oliveira}},
  \bibinfo{author}{\bibfnamefont{L.}~\bibnamefont{Reining}}, \bibnamefont{and}
  \bibinfo{author}{\bibfnamefont{S.}~\bibnamefont{Botti}},
  \bibinfo{journal}{Phys. Rev. B} \textbf{\bibinfo{volume}{83}},
  \bibinfo{pages}{035119} (\bibinfo{year}{2011}).

\bibitem[{\citenamefont{Skone et~al.}(2014)\citenamefont{Skone, Govoni, and
  Galli}}]{skone2014self}
\bibinfo{author}{\bibfnamefont{J.~H.} \bibnamefont{Skone}},
  \bibinfo{author}{\bibfnamefont{M.}~\bibnamefont{Govoni}}, \bibnamefont{and}
  \bibinfo{author}{\bibfnamefont{G.}~\bibnamefont{Galli}},
  \bibinfo{journal}{Phys. Rev. B} \textbf{\bibinfo{volume}{89}},
  \bibinfo{pages}{195112} (\bibinfo{year}{2014}).

\bibitem[{\citenamefont{Erba}(2017)}]{MIOSCHYB}
\bibinfo{author}{\bibfnamefont{A.}~\bibnamefont{Erba}}, \bibinfo{journal}{J.
  Phys.: Condens. Matter} \textbf{\bibinfo{volume}{29}},
  \bibinfo{pages}{314001} (\bibinfo{year}{2017}).

\bibitem[{\citenamefont{Kronik et~al.}(2012)\citenamefont{Kronik, Stein,
  Refaely-Abramson, and Baer}}]{Kronik2012}
\bibinfo{author}{\bibfnamefont{L.}~\bibnamefont{Kronik}},
  \bibinfo{author}{\bibfnamefont{T.}~\bibnamefont{Stein}},
  \bibinfo{author}{\bibfnamefont{S.}~\bibnamefont{Refaely-Abramson}},
  \bibnamefont{and} \bibinfo{author}{\bibfnamefont{R.}~\bibnamefont{Baer}},
  \bibinfo{journal}{J. Chem. Theory Comput.} \textbf{\bibinfo{volume}{8}},
  \bibinfo{pages}{1515} (\bibinfo{year}{2012}).

\bibitem[{\citenamefont{Nguyen et~al.}(2018)\citenamefont{Nguyen, Colonna,
  Ferretti, and Marzari}}]{Nguyen2018}
\bibinfo{author}{\bibfnamefont{N.~L.} \bibnamefont{Nguyen}},
  \bibinfo{author}{\bibfnamefont{N.}~\bibnamefont{Colonna}},
  \bibinfo{author}{\bibfnamefont{A.}~\bibnamefont{Ferretti}}, \bibnamefont{and}
  \bibinfo{author}{\bibfnamefont{N.}~\bibnamefont{Marzari}},
  \bibinfo{journal}{Phys. Rev. X} \textbf{\bibinfo{volume}{8}},
  \bibinfo{pages}{021051} (\bibinfo{year}{2018}).

\bibitem[{\citenamefont{Prokopiou et~al.}(2022)\citenamefont{Prokopiou,
  Hartstein, Govind, and Kronik}}]{Prokopiou2022}
\bibinfo{author}{\bibfnamefont{G.}~\bibnamefont{Prokopiou}},
  \bibinfo{author}{\bibfnamefont{M.}~\bibnamefont{Hartstein}},
  \bibinfo{author}{\bibfnamefont{N.}~\bibnamefont{Govind}}, \bibnamefont{and}
  \bibinfo{author}{\bibfnamefont{L.}~\bibnamefont{Kronik}},
  \bibinfo{journal}{J. Chem. Theory Comput.} \textbf{\bibinfo{volume}{18}},
  \bibinfo{pages}{2331} (\bibinfo{year}{2022}).

\bibitem[{\citenamefont{Peralta et~al.}(2007)\citenamefont{Peralta, Scuseria,
  and Frisch}}]{Peralta2007}
\bibinfo{author}{\bibfnamefont{J.~E.} \bibnamefont{Peralta}},
  \bibinfo{author}{\bibfnamefont{G.~E.} \bibnamefont{Scuseria}},
  \bibnamefont{and} \bibinfo{author}{\bibfnamefont{M.~J.}
  \bibnamefont{Frisch}}, \bibinfo{journal}{Phys. Rev. B}
  \textbf{\bibinfo{volume}{75}}, \bibinfo{pages}{125119}
  (\bibinfo{year}{2007}),
  \urlprefix\url{https://link.aps.org/doi/10.1103/PhysRevB.75.125119}.

\bibitem[{\citenamefont{Huhn and Blum}(2017)}]{Huhn2017}
\bibinfo{author}{\bibfnamefont{W.~P.} \bibnamefont{Huhn}} \bibnamefont{and}
  \bibinfo{author}{\bibfnamefont{V.}~\bibnamefont{Blum}},
  \bibinfo{journal}{Phys. Rev. Mater.} \textbf{\bibinfo{volume}{1}},
  \bibinfo{pages}{033803} (\bibinfo{year}{2017}),
  \urlprefix\url{https://link.aps.org/doi/10.1103/PhysRevMaterials.1.033803}.

\bibitem[{\citenamefont{Koelling and Harmon}(1977)}]{koelling1977technique}
\bibinfo{author}{\bibfnamefont{D.}~\bibnamefont{Koelling}} \bibnamefont{and}
  \bibinfo{author}{\bibfnamefont{B.}~\bibnamefont{Harmon}},
  \bibinfo{journal}{J. Phys. C Sol. State Phys.} \textbf{\bibinfo{volume}{10}},
  \bibinfo{pages}{3107} (\bibinfo{year}{1977}).

\bibitem[{\citenamefont{Desmarais
  et~al.}(2021{\natexlab{b}})\citenamefont{Desmarais, Erba, Flament, and
  Kirtman}}]{desmarais2021perturbation}
\bibinfo{author}{\bibfnamefont{J.~K.} \bibnamefont{Desmarais}},
  \bibinfo{author}{\bibfnamefont{A.}~\bibnamefont{Erba}},
  \bibinfo{author}{\bibfnamefont{J.-P.} \bibnamefont{Flament}},
  \bibnamefont{and} \bibinfo{author}{\bibfnamefont{B.}~\bibnamefont{Kirtman}},
  \bibinfo{journal}{J. Chem. Theor. Comput.} \textbf{\bibinfo{volume}{17}},
  \bibinfo{pages}{4712} (\bibinfo{year}{2021}{\natexlab{b}}).

\bibitem[{\citenamefont{Desmarais et~al.}(2023)\citenamefont{Desmarais,
  Boccuni, Flament, Kirtman, and Erba}}]{desmarais2023perturbation}
\bibinfo{author}{\bibfnamefont{J.~K.} \bibnamefont{Desmarais}},
  \bibinfo{author}{\bibfnamefont{A.}~\bibnamefont{Boccuni}},
  \bibinfo{author}{\bibfnamefont{J.-P.} \bibnamefont{Flament}},
  \bibinfo{author}{\bibfnamefont{B.}~\bibnamefont{Kirtman}}, \bibnamefont{and}
  \bibinfo{author}{\bibfnamefont{A.}~\bibnamefont{Erba}}, \bibinfo{journal}{J.
  Chem. Theor. Comput.} \textbf{\bibinfo{volume}{19}}, \bibinfo{pages}{1853}
  (\bibinfo{year}{2023}).

\bibitem[{\citenamefont{Lee and Parr}(1987)}]{LP87}
\bibinfo{author}{\bibfnamefont{C.}~\bibnamefont{Lee}} \bibnamefont{and}
  \bibinfo{author}{\bibfnamefont{R.~G.} \bibnamefont{Parr}},
  \bibinfo{journal}{Phys. Rev. A} \textbf{\bibinfo{volume}{35}},
  \bibinfo{pages}{2377} (\bibinfo{year}{1987}).

\bibitem[{\citenamefont{Dobson}(1993)}]{Dobson93}
\bibinfo{author}{\bibfnamefont{J.~F.} \bibnamefont{Dobson}},
  \bibinfo{journal}{The Journal of Chemical Physics}
  \textbf{\bibinfo{volume}{98}}, \bibinfo{pages}{8870} (\bibinfo{year}{1993}).

\bibitem[{\citenamefont{Becke}(1996)}]{Becke96}
\bibinfo{author}{\bibfnamefont{A.~D.} \bibnamefont{Becke}},
  \bibinfo{journal}{Canadian Journal of Chemistry}
  \textbf{\bibinfo{volume}{74}}, \bibinfo{pages}{995} (\bibinfo{year}{1996}).

\bibitem[{\citenamefont{Becke}(2002)}]{Becke-j02}
\bibinfo{author}{\bibfnamefont{A.~D.} \bibnamefont{Becke}},
  \bibinfo{journal}{The Journal of Chemical Physics}
  \textbf{\bibinfo{volume}{117}}, \bibinfo{pages}{6935} (\bibinfo{year}{2002}).

\bibitem[{\citenamefont{Maximoff et~al.}(2004)\citenamefont{Maximoff,
  Ernzerhof, and Scuseria}}]{J-PBE}
\bibinfo{author}{\bibfnamefont{S.~N.} \bibnamefont{Maximoff}},
  \bibinfo{author}{\bibfnamefont{M.}~\bibnamefont{Ernzerhof}},
  \bibnamefont{and} \bibinfo{author}{\bibfnamefont{G.~E.}
  \bibnamefont{Scuseria}}, \bibinfo{journal}{The Journal of Chemical Physics}
  \textbf{\bibinfo{volume}{120}}, \bibinfo{pages}{2105} (\bibinfo{year}{2004}).

\bibitem[{\citenamefont{Burnus et~al.}(2005)\citenamefont{Burnus, Marques, and
  Gross}}]{burnus2005time}
\bibinfo{author}{\bibfnamefont{T.}~\bibnamefont{Burnus}},
  \bibinfo{author}{\bibfnamefont{M.~A.} \bibnamefont{Marques}},
  \bibnamefont{and} \bibinfo{author}{\bibfnamefont{E.~K.} \bibnamefont{Gross}},
  \bibinfo{journal}{Physical Review A} \textbf{\bibinfo{volume}{71}},
  \bibinfo{pages}{010501} (\bibinfo{year}{2005}).

\bibitem[{\citenamefont{Tao and Perdew}(2005)}]{TP05}
\bibinfo{author}{\bibfnamefont{J.}~\bibnamefont{Tao}} \bibnamefont{and}
  \bibinfo{author}{\bibfnamefont{J.~P.} \bibnamefont{Perdew}},
  \bibinfo{journal}{Phys. Rev. Lett.} \textbf{\bibinfo{volume}{95}},
  \bibinfo{pages}{196403} (\bibinfo{year}{2005}).

\bibitem[{\citenamefont{Pittalis et~al.}(2007)\citenamefont{Pittalis, Kurth,
  Sharma, and Gross}}]{Pittalis07}
\bibinfo{author}{\bibfnamefont{S.}~\bibnamefont{Pittalis}},
  \bibinfo{author}{\bibfnamefont{S.}~\bibnamefont{Kurth}},
  \bibinfo{author}{\bibfnamefont{S.}~\bibnamefont{Sharma}}, \bibnamefont{and}
  \bibinfo{author}{\bibfnamefont{E.~K.~U.} \bibnamefont{Gross}},
  \bibinfo{journal}{The Journal of Chemical Physics}
  \textbf{\bibinfo{volume}{127}}, \bibinfo{pages}{124103}
  (\bibinfo{year}{2007}).

\bibitem[{\citenamefont{Pittalis et~al.}(2009)\citenamefont{Pittalis,
  R\"as\"anen, and Gross}}]{Pittalis09}
\bibinfo{author}{\bibfnamefont{S.}~\bibnamefont{Pittalis}},
  \bibinfo{author}{\bibfnamefont{E.}~\bibnamefont{R\"as\"anen}},
  \bibnamefont{and} \bibinfo{author}{\bibfnamefont{E.~K.~U.}
  \bibnamefont{Gross}}, \bibinfo{journal}{Phys. Rev. A}
  \textbf{\bibinfo{volume}{80}}, \bibinfo{pages}{032515}
  (\bibinfo{year}{2009}).

\bibitem[{\citenamefont{R\"as\"anen et~al.}(2009)\citenamefont{R\"as\"anen,
  Pittalis, Proetto, and Gross}}]{Rasanen09}
\bibinfo{author}{\bibfnamefont{E.}~\bibnamefont{R\"as\"anen}},
  \bibinfo{author}{\bibfnamefont{S.}~\bibnamefont{Pittalis}},
  \bibinfo{author}{\bibfnamefont{C.~R.} \bibnamefont{Proetto}},
  \bibnamefont{and} \bibinfo{author}{\bibfnamefont{E.~K.~U.}
  \bibnamefont{Gross}}, \bibinfo{journal}{Phys. Rev. B}
  \textbf{\bibinfo{volume}{79}}, \bibinfo{pages}{121305}
  (\bibinfo{year}{2009}).

\bibitem[{\citenamefont{Oliveira et~al.}(2010)\citenamefont{Oliveira,
  Räsänen, Pittalis, and Marques}}]{Oliveira2010}
\bibinfo{author}{\bibfnamefont{M.~J.~T.} \bibnamefont{Oliveira}},
  \bibinfo{author}{\bibfnamefont{E.}~\bibnamefont{Räsänen}},
  \bibinfo{author}{\bibfnamefont{S.}~\bibnamefont{Pittalis}}, \bibnamefont{and}
  \bibinfo{author}{\bibfnamefont{M.~A.~L.} \bibnamefont{Marques}},
  \bibinfo{journal}{Journal of Chemical Theory and Computation}
  \textbf{\bibinfo{volume}{6}}, \bibinfo{pages}{3664} (\bibinfo{year}{2010}).

\bibitem[{\citenamefont{Zhu et~al.}(2016)\citenamefont{Zhu, Zhang, and
  Trickey}}]{Tricky16}
\bibinfo{author}{\bibfnamefont{W.}~\bibnamefont{Zhu}},
  \bibinfo{author}{\bibfnamefont{L.}~\bibnamefont{Zhang}}, \bibnamefont{and}
  \bibinfo{author}{\bibfnamefont{S.~B.} \bibnamefont{Trickey}},
  \bibinfo{journal}{The Journal of Chemical Physics}
  \textbf{\bibinfo{volume}{145}}, \bibinfo{pages}{224106}
  (\bibinfo{year}{2016}),
  \eprint{http://aip.scitation.org/doi/pdf/10.1063/1.4971377},
  \urlprefix\url{http://aip.scitation.org/doi/abs/10.1063/1.4971377}.

\bibitem[{\citenamefont{James W.~Furness and Teale}(2016)}]{Furness2016}
\bibinfo{author}{\bibfnamefont{T.~H.} \bibnamefont{James W.~Furness},
  \bibfnamefont{Ulf~Ekström}} \bibnamefont{and}
  \bibinfo{author}{\bibfnamefont{A.~M.} \bibnamefont{Teale}},
  \bibinfo{journal}{Molecular Physics} \textbf{\bibinfo{volume}{114}},
  \bibinfo{pages}{1415} (\bibinfo{year}{2016}).

\bibitem[{\citenamefont{Holzer et~al.}(2022)\citenamefont{Holzer, Franzke, and
  Pausch}}]{Holzer2022}
\bibinfo{author}{\bibfnamefont{C.}~\bibnamefont{Holzer}},
  \bibinfo{author}{\bibfnamefont{Y.~J.} \bibnamefont{Franzke}},
  \bibnamefont{and} \bibinfo{author}{\bibfnamefont{A.}~\bibnamefont{Pausch}},
  \bibinfo{journal}{The Journal of Chemical Physics}
  \textbf{\bibinfo{volume}{157}}, \bibinfo{pages}{204102}
  (\bibinfo{year}{2022}), ISSN \bibinfo{issn}{0021-9606}.

\bibitem[{\citenamefont{Camino et~al.}(2023)\citenamefont{Camino, Zhou,
  Ascrizzi, Boccuni, Bodo, Cossard, Mitoli, Ferrari, Erba, and
  Harrison}}]{CRYSTALpytools}
\bibinfo{author}{\bibfnamefont{B.}~\bibnamefont{Camino}},
  \bibinfo{author}{\bibfnamefont{H.}~\bibnamefont{Zhou}},
  \bibinfo{author}{\bibfnamefont{E.}~\bibnamefont{Ascrizzi}},
  \bibinfo{author}{\bibfnamefont{A.}~\bibnamefont{Boccuni}},
  \bibinfo{author}{\bibfnamefont{F.}~\bibnamefont{Bodo}},
  \bibinfo{author}{\bibfnamefont{A.}~\bibnamefont{Cossard}},
  \bibinfo{author}{\bibfnamefont{D.}~\bibnamefont{Mitoli}},
  \bibinfo{author}{\bibfnamefont{A.~M.} \bibnamefont{Ferrari}},
  \bibinfo{author}{\bibfnamefont{A.}~\bibnamefont{Erba}}, \bibnamefont{and}
  \bibinfo{author}{\bibfnamefont{N.~M.} \bibnamefont{Harrison}},
  \bibinfo{journal}{Comput. Phys. Commun.} \textbf{\bibinfo{volume}{292}},
  \bibinfo{pages}{108853} (\bibinfo{year}{2023}).

\bibitem[{\citenamefont{Rashba}(1960)}]{rashba1960properties}
\bibinfo{author}{\bibfnamefont{E.}~\bibnamefont{Rashba}},
  \bibinfo{journal}{Sov. Phys.-Solid State} \textbf{\bibinfo{volume}{2}},
  \bibinfo{pages}{1109} (\bibinfo{year}{1960}).

\bibitem[{\citenamefont{Zhang et~al.}(2014{\natexlab{a}})\citenamefont{Zhang,
  Liu, Luo, Freeman, and Zunger}}]{zhang2014hidden}
\bibinfo{author}{\bibfnamefont{X.}~\bibnamefont{Zhang}},
  \bibinfo{author}{\bibfnamefont{Q.}~\bibnamefont{Liu}},
  \bibinfo{author}{\bibfnamefont{J.-W.} \bibnamefont{Luo}},
  \bibinfo{author}{\bibfnamefont{A.~J.} \bibnamefont{Freeman}},
  \bibnamefont{and} \bibinfo{author}{\bibfnamefont{A.}~\bibnamefont{Zunger}},
  \bibinfo{journal}{Nat. Phys.} \textbf{\bibinfo{volume}{10}},
  \bibinfo{pages}{387} (\bibinfo{year}{2014}{\natexlab{a}}).

\bibitem[{\citenamefont{Oliva et~al.}(2020)\citenamefont{Oliva, Wo{\'z}niak,
  Dybala, Kopaczek, Scharoch, and Kudrawiec}}]{oliva2020hidden}
\bibinfo{author}{\bibfnamefont{R.}~\bibnamefont{Oliva}},
  \bibinfo{author}{\bibfnamefont{T.}~\bibnamefont{Wo{\'z}niak}},
  \bibinfo{author}{\bibfnamefont{F.}~\bibnamefont{Dybala}},
  \bibinfo{author}{\bibfnamefont{J.}~\bibnamefont{Kopaczek}},
  \bibinfo{author}{\bibfnamefont{P.}~\bibnamefont{Scharoch}}, \bibnamefont{and}
  \bibinfo{author}{\bibfnamefont{R.}~\bibnamefont{Kudrawiec}},
  \bibinfo{journal}{Mat. Res. Lett.} \textbf{\bibinfo{volume}{8}},
  \bibinfo{pages}{75} (\bibinfo{year}{2020}).

\bibitem[{\citenamefont{Zhang et~al.}(2014{\natexlab{b}})\citenamefont{Zhang,
  Chang, Zhou, Cui, Yan, Liu, Schmitt, Lee, Moore, Chen
  et~al.}}]{zhang2014direct}
\bibinfo{author}{\bibfnamefont{Y.}~\bibnamefont{Zhang}},
  \bibinfo{author}{\bibfnamefont{T.-R.} \bibnamefont{Chang}},
  \bibinfo{author}{\bibfnamefont{B.}~\bibnamefont{Zhou}},
  \bibinfo{author}{\bibfnamefont{Y.-T.} \bibnamefont{Cui}},
  \bibinfo{author}{\bibfnamefont{H.}~\bibnamefont{Yan}},
  \bibinfo{author}{\bibfnamefont{Z.}~\bibnamefont{Liu}},
  \bibinfo{author}{\bibfnamefont{F.}~\bibnamefont{Schmitt}},
  \bibinfo{author}{\bibfnamefont{J.}~\bibnamefont{Lee}},
  \bibinfo{author}{\bibfnamefont{R.}~\bibnamefont{Moore}},
  \bibinfo{author}{\bibfnamefont{Y.}~\bibnamefont{Chen}}, \bibnamefont{et~al.},
  \bibinfo{journal}{Nature nano.} \textbf{\bibinfo{volume}{9}},
  \bibinfo{pages}{111} (\bibinfo{year}{2014}{\natexlab{b}}).

\bibitem[{\citenamefont{Shim et~al.}(2014)\citenamefont{Shim, Yoo, Seo, Shin,
  Jung, Kang, Ahn, Cho, and Choi}}]{shim2014large}
\bibinfo{author}{\bibfnamefont{G.~W.} \bibnamefont{Shim}},
  \bibinfo{author}{\bibfnamefont{K.}~\bibnamefont{Yoo}},
  \bibinfo{author}{\bibfnamefont{S.-B.} \bibnamefont{Seo}},
  \bibinfo{author}{\bibfnamefont{J.}~\bibnamefont{Shin}},
  \bibinfo{author}{\bibfnamefont{D.~Y.} \bibnamefont{Jung}},
  \bibinfo{author}{\bibfnamefont{I.-S.} \bibnamefont{Kang}},
  \bibinfo{author}{\bibfnamefont{C.~W.} \bibnamefont{Ahn}},
  \bibinfo{author}{\bibfnamefont{B.~J.} \bibnamefont{Cho}}, \bibnamefont{and}
  \bibinfo{author}{\bibfnamefont{S.-Y.} \bibnamefont{Choi}},
  \bibinfo{journal}{ACS nano} \textbf{\bibinfo{volume}{8}},
  \bibinfo{pages}{6655} (\bibinfo{year}{2014}).

\bibitem[{\citenamefont{Ross et~al.}(2013)\citenamefont{Ross, Wu, Yu, Ghimire,
  Jones, Aivazian, Yan, Mandrus, Xiao, Yao et~al.}}]{ross2013electrical}
\bibinfo{author}{\bibfnamefont{J.~S.} \bibnamefont{Ross}},
  \bibinfo{author}{\bibfnamefont{S.}~\bibnamefont{Wu}},
  \bibinfo{author}{\bibfnamefont{H.}~\bibnamefont{Yu}},
  \bibinfo{author}{\bibfnamefont{N.~J.} \bibnamefont{Ghimire}},
  \bibinfo{author}{\bibfnamefont{A.~M.} \bibnamefont{Jones}},
  \bibinfo{author}{\bibfnamefont{G.}~\bibnamefont{Aivazian}},
  \bibinfo{author}{\bibfnamefont{J.}~\bibnamefont{Yan}},
  \bibinfo{author}{\bibfnamefont{D.~G.} \bibnamefont{Mandrus}},
  \bibinfo{author}{\bibfnamefont{D.}~\bibnamefont{Xiao}},
  \bibinfo{author}{\bibfnamefont{W.}~\bibnamefont{Yao}}, \bibnamefont{et~al.},
  \bibinfo{journal}{Nature Commun.} \textbf{\bibinfo{volume}{4}},
  \bibinfo{pages}{1474} (\bibinfo{year}{2013}).

\bibitem[{\citenamefont{Erba et~al.}(2022)\citenamefont{Erba, Desmarais,
  Casassa, Civalleri, Donà, Bush, Searle, Maschio, Edith-Daga, Cossard
  et~al.}}]{Erba2023}
\bibinfo{author}{\bibfnamefont{A.}~\bibnamefont{Erba}},
  \bibinfo{author}{\bibfnamefont{J.~K.} \bibnamefont{Desmarais}},
  \bibinfo{author}{\bibfnamefont{S.}~\bibnamefont{Casassa}},
  \bibinfo{author}{\bibfnamefont{B.}~\bibnamefont{Civalleri}},
  \bibinfo{author}{\bibfnamefont{L.}~\bibnamefont{Donà}},
  \bibinfo{author}{\bibfnamefont{I.~J.} \bibnamefont{Bush}},
  \bibinfo{author}{\bibfnamefont{B.}~\bibnamefont{Searle}},
  \bibinfo{author}{\bibfnamefont{L.}~\bibnamefont{Maschio}},
  \bibinfo{author}{\bibfnamefont{L.}~\bibnamefont{Edith-Daga}},
  \bibinfo{author}{\bibfnamefont{A.}~\bibnamefont{Cossard}},
  \bibnamefont{et~al.}, \bibinfo{journal}{J. Chem. Theory Comput.}
  (\bibinfo{year}{2022}), \urlprefix\url{10.1021/acs.jctc.2c00958}.

\bibitem[{\citenamefont{Perdew et~al.}(1996{\natexlab{b}})\citenamefont{Perdew,
  Burke, and Ernzerhof}}]{perdew1996generalized}
\bibinfo{author}{\bibfnamefont{J.~P.} \bibnamefont{Perdew}},
  \bibinfo{author}{\bibfnamefont{K.}~\bibnamefont{Burke}}, \bibnamefont{and}
  \bibinfo{author}{\bibfnamefont{M.}~\bibnamefont{Ernzerhof}},
  \bibinfo{journal}{Physical review letters} \textbf{\bibinfo{volume}{77}},
  \bibinfo{pages}{3865} (\bibinfo{year}{1996}{\natexlab{b}}).

\bibitem[{ESI()}]{ESI}
\bibinfo{note}{Computational details are available at INSERT URL HERE}.

\bibitem[{\citenamefont{Doll et~al.}(2001)\citenamefont{Doll, Saunders, and
  Harrison}}]{doll2001analytical}
\bibinfo{author}{\bibfnamefont{K.}~\bibnamefont{Doll}},
  \bibinfo{author}{\bibfnamefont{V.}~\bibnamefont{Saunders}}, \bibnamefont{and}
  \bibinfo{author}{\bibfnamefont{N.}~\bibnamefont{Harrison}},
  \bibinfo{journal}{Int. J. Q. Chem.} \textbf{\bibinfo{volume}{82}},
  \bibinfo{pages}{1} (\bibinfo{year}{2001}).

\bibitem[{\citenamefont{Doll}(2001)}]{doll2001implementation}
\bibinfo{author}{\bibfnamefont{K.}~\bibnamefont{Doll}},
  \bibinfo{journal}{Comput. Phys. Commun.} \textbf{\bibinfo{volume}{137}},
  \bibinfo{pages}{74} (\bibinfo{year}{2001}).

\bibitem[{\citenamefont{Doll et~al.}(2006)\citenamefont{Doll, Dovesi, and
  Orlando}}]{doll2006analytical}
\bibinfo{author}{\bibfnamefont{K.}~\bibnamefont{Doll}},
  \bibinfo{author}{\bibfnamefont{R.}~\bibnamefont{Dovesi}}, \bibnamefont{and}
  \bibinfo{author}{\bibfnamefont{R.}~\bibnamefont{Orlando}},
  \bibinfo{journal}{Theor. Chem. Acc.} \textbf{\bibinfo{volume}{115}},
  \bibinfo{pages}{354} (\bibinfo{year}{2006}).

\bibitem[{\citenamefont{Civalleri et~al.}(2001)\citenamefont{Civalleri, D'Arco,
  Orlando, Saunders, and Dovesi}}]{civalleri2001hartree}
\bibinfo{author}{\bibfnamefont{B.}~\bibnamefont{Civalleri}},
  \bibinfo{author}{\bibfnamefont{P.}~\bibnamefont{D'Arco}},
  \bibinfo{author}{\bibfnamefont{R.}~\bibnamefont{Orlando}},
  \bibinfo{author}{\bibfnamefont{V.}~\bibnamefont{Saunders}}, \bibnamefont{and}
  \bibinfo{author}{\bibfnamefont{R.}~\bibnamefont{Dovesi}},
  \bibinfo{journal}{Chem. Phys. Lett.} \textbf{\bibinfo{volume}{348}},
  \bibinfo{pages}{131} (\bibinfo{year}{2001}).

\bibitem[{\citenamefont{Metz et~al.}(2000)\citenamefont{Metz, Stoll, and
  Dolg}}]{metz2000small}
\bibinfo{author}{\bibfnamefont{B.}~\bibnamefont{Metz}},
  \bibinfo{author}{\bibfnamefont{H.}~\bibnamefont{Stoll}}, \bibnamefont{and}
  \bibinfo{author}{\bibfnamefont{M.}~\bibnamefont{Dolg}}, \bibinfo{journal}{J.
  Chem. Phys.} \textbf{\bibinfo{volume}{113}}, \bibinfo{pages}{2563}
  (\bibinfo{year}{2000}).

\bibitem[{\citenamefont{Peterson et~al.}(2003)\citenamefont{Peterson, Figgen,
  Goll, Stoll, and Dolg}}]{peterson2003systematically}
\bibinfo{author}{\bibfnamefont{K.~A.} \bibnamefont{Peterson}},
  \bibinfo{author}{\bibfnamefont{D.}~\bibnamefont{Figgen}},
  \bibinfo{author}{\bibfnamefont{E.}~\bibnamefont{Goll}},
  \bibinfo{author}{\bibfnamefont{H.}~\bibnamefont{Stoll}}, \bibnamefont{and}
  \bibinfo{author}{\bibfnamefont{M.}~\bibnamefont{Dolg}}, \bibinfo{journal}{J.
  Chem. Phys.} \textbf{\bibinfo{volume}{119}}, \bibinfo{pages}{11113}
  (\bibinfo{year}{2003}).

\bibitem[{\citenamefont{Peterson et~al.}(2007)\citenamefont{Peterson, Figgen,
  Dolg, and Stoll}}]{peterson2007energy}
\bibinfo{author}{\bibfnamefont{K.~A.} \bibnamefont{Peterson}},
  \bibinfo{author}{\bibfnamefont{D.}~\bibnamefont{Figgen}},
  \bibinfo{author}{\bibfnamefont{M.}~\bibnamefont{Dolg}}, \bibnamefont{and}
  \bibinfo{author}{\bibfnamefont{H.}~\bibnamefont{Stoll}}, \bibinfo{journal}{J.
  Chem. Phys.} \textbf{\bibinfo{volume}{126}} (\bibinfo{year}{2007}).

\bibitem[{\citenamefont{Laun et~al.}(2018)\citenamefont{Laun, Vilela~Oliveira,
  and Bredow}}]{laun2018consistent}
\bibinfo{author}{\bibfnamefont{J.}~\bibnamefont{Laun}},
  \bibinfo{author}{\bibfnamefont{D.}~\bibnamefont{Vilela~Oliveira}},
  \bibnamefont{and} \bibinfo{author}{\bibfnamefont{T.}~\bibnamefont{Bredow}},
  \bibinfo{journal}{J. Comp. Chem.} \textbf{\bibinfo{volume}{39}},
  \bibinfo{pages}{1285} (\bibinfo{year}{2018}).

\bibitem[{\citenamefont{Heyd et~al.}(2005)\citenamefont{Heyd, Peralta,
  Scuseria, and Martin}}]{heyd2005energy}
\bibinfo{author}{\bibfnamefont{J.}~\bibnamefont{Heyd}},
  \bibinfo{author}{\bibfnamefont{J.~E.} \bibnamefont{Peralta}},
  \bibinfo{author}{\bibfnamefont{G.~E.} \bibnamefont{Scuseria}},
  \bibnamefont{and} \bibinfo{author}{\bibfnamefont{R.~L.}
  \bibnamefont{Martin}}, \bibinfo{journal}{J. Chem. Phys.}
  \textbf{\bibinfo{volume}{123}} (\bibinfo{year}{2005}).

\bibitem[{\citenamefont{Lebedev}(1976)}]{lebedev1976quadratures}
\bibinfo{author}{\bibfnamefont{V.~I.} \bibnamefont{Lebedev}},
  \bibinfo{journal}{USSR Comput. Math. Math. Phys.}
  \textbf{\bibinfo{volume}{16}}, \bibinfo{pages}{10} (\bibinfo{year}{1976}).

\bibitem[{\citenamefont{Lebedev}(1977)}]{lebedev1977spherical}
\bibinfo{author}{\bibfnamefont{V.~I.} \bibnamefont{Lebedev}},
  \bibinfo{journal}{Sib. Math. J.} \textbf{\bibinfo{volume}{18}},
  \bibinfo{pages}{99} (\bibinfo{year}{1977}).

\bibitem[{\citenamefont{Towler et~al.}(1996)\citenamefont{Towler, Zupan, and
  Caus{\`a}}}]{towler1996density}
\bibinfo{author}{\bibfnamefont{M.~D.} \bibnamefont{Towler}},
  \bibinfo{author}{\bibfnamefont{A.}~\bibnamefont{Zupan}}, \bibnamefont{and}
  \bibinfo{author}{\bibfnamefont{M.}~\bibnamefont{Caus{\`a}}},
  \bibinfo{journal}{Comput. Phys. Commun.} \textbf{\bibinfo{volume}{98}},
  \bibinfo{pages}{181} (\bibinfo{year}{1996}).

\bibitem[{\citenamefont{Beal et~al.}(1972)\citenamefont{Beal, Knights, and
  Liang}}]{beal1972transmission}
\bibinfo{author}{\bibfnamefont{A.}~\bibnamefont{Beal}},
  \bibinfo{author}{\bibfnamefont{J.}~\bibnamefont{Knights}}, \bibnamefont{and}
  \bibinfo{author}{\bibfnamefont{W.}~\bibnamefont{Liang}}, \bibinfo{journal}{J.
  Phys. C Sol. State Phys.} \textbf{\bibinfo{volume}{5}}, \bibinfo{pages}{3540}
  (\bibinfo{year}{1972}).

\bibitem[{\citenamefont{Ruppert
  et~al.}(2014{\natexlab{a}})\citenamefont{Ruppert, Aslan, and
  Heinz}}]{ruppert2014optical}
\bibinfo{author}{\bibfnamefont{C.}~\bibnamefont{Ruppert}},
  \bibinfo{author}{\bibfnamefont{B.}~\bibnamefont{Aslan}}, \bibnamefont{and}
  \bibinfo{author}{\bibfnamefont{T.~F.} \bibnamefont{Heinz}},
  \bibinfo{journal}{Nano Lett.} \textbf{\bibinfo{volume}{14}},
  \bibinfo{pages}{6231} (\bibinfo{year}{2014}{\natexlab{a}}).

\bibitem[{\citenamefont{Choi et~al.}(2017)\citenamefont{Choi, Kim, Jung, Kim,
  Yu, and Chang}}]{choi2017temperature}
\bibinfo{author}{\bibfnamefont{B.~K.} \bibnamefont{Choi}},
  \bibinfo{author}{\bibfnamefont{M.}~\bibnamefont{Kim}},
  \bibinfo{author}{\bibfnamefont{K.-H.} \bibnamefont{Jung}},
  \bibinfo{author}{\bibfnamefont{J.}~\bibnamefont{Kim}},
  \bibinfo{author}{\bibfnamefont{K.-S.} \bibnamefont{Yu}}, \bibnamefont{and}
  \bibinfo{author}{\bibfnamefont{Y.~J.} \bibnamefont{Chang}},
  \bibinfo{journal}{Nanoscale Res. Lett.} \textbf{\bibinfo{volume}{12}},
  \bibinfo{pages}{1} (\bibinfo{year}{2017}).

\bibitem[{\citenamefont{Island et~al.}(2016)\citenamefont{Island, Kuc,
  Diependaal, Bratschitsch, Van Der~Zant, Heine, and
  Castellanos-Gomez}}]{island2016precise}
\bibinfo{author}{\bibfnamefont{J.~O.} \bibnamefont{Island}},
  \bibinfo{author}{\bibfnamefont{A.}~\bibnamefont{Kuc}},
  \bibinfo{author}{\bibfnamefont{E.~H.} \bibnamefont{Diependaal}},
  \bibinfo{author}{\bibfnamefont{R.}~\bibnamefont{Bratschitsch}},
  \bibinfo{author}{\bibfnamefont{H.~S.} \bibnamefont{Van Der~Zant}},
  \bibinfo{author}{\bibfnamefont{T.}~\bibnamefont{Heine}}, \bibnamefont{and}
  \bibinfo{author}{\bibfnamefont{A.}~\bibnamefont{Castellanos-Gomez}},
  \bibinfo{journal}{Nanoscale} \textbf{\bibinfo{volume}{8}},
  \bibinfo{pages}{2589} (\bibinfo{year}{2016}).

\bibitem[{\citenamefont{Zelewski and
  Kudrawiec}(2017)}]{zelewski2017photoacoustic}
\bibinfo{author}{\bibfnamefont{S.~J.} \bibnamefont{Zelewski}} \bibnamefont{and}
  \bibinfo{author}{\bibfnamefont{R.}~\bibnamefont{Kudrawiec}},
  \bibinfo{journal}{Sci. Reports} \textbf{\bibinfo{volume}{7}},
  \bibinfo{pages}{15365} (\bibinfo{year}{2017}).

\bibitem[{\citenamefont{Ruppert
  et~al.}(2014{\natexlab{b}})\citenamefont{Ruppert, Aslan, and
  Heinz}}]{ruppert}
\bibinfo{author}{\bibfnamefont{C.}~\bibnamefont{Ruppert}},
  \bibinfo{author}{\bibfnamefont{B.}~\bibnamefont{Aslan}}, \bibnamefont{and}
  \bibinfo{author}{\bibfnamefont{T.~F.} \bibnamefont{Heinz}},
  \bibinfo{journal}{Nano Lett.} \textbf{\bibinfo{volume}{14}},
  \bibinfo{pages}{6231} (\bibinfo{year}{2014}{\natexlab{b}}).

\bibitem[{\citenamefont{Conan et~al.}(1979)\citenamefont{Conan, Delaunay,
  Bonnet, Moustafa, and Spiesser}}]{conan1979temperature}
\bibinfo{author}{\bibfnamefont{A.}~\bibnamefont{Conan}},
  \bibinfo{author}{\bibfnamefont{D.}~\bibnamefont{Delaunay}},
  \bibinfo{author}{\bibfnamefont{A.}~\bibnamefont{Bonnet}},
  \bibinfo{author}{\bibfnamefont{A.}~\bibnamefont{Moustafa}}, \bibnamefont{and}
  \bibinfo{author}{\bibfnamefont{M.}~\bibnamefont{Spiesser}},
  \bibinfo{journal}{physica status solidi (b)} \textbf{\bibinfo{volume}{94}},
  \bibinfo{pages}{279} (\bibinfo{year}{1979}).

\bibitem[{\citenamefont{Sun et~al.}(2015)\citenamefont{Sun, Ruzsinszky, and
  Perdew}}]{SCAN}
\bibinfo{author}{\bibfnamefont{J.}~\bibnamefont{Sun}},
  \bibinfo{author}{\bibfnamefont{A.}~\bibnamefont{Ruzsinszky}},
  \bibnamefont{and} \bibinfo{author}{\bibfnamefont{J.~P.}
  \bibnamefont{Perdew}}, \bibinfo{journal}{Phys. Rev. Lett.}
  \textbf{\bibinfo{volume}{115}}, \bibinfo{pages}{036402}
  (\bibinfo{year}{2015}),
  \urlprefix\url{https://link.aps.org/doi/10.1103/PhysRevLett.115.036402}.

\bibitem[{\citenamefont{Aschebrock and K\"ummel}(2019)}]{TASK}
\bibinfo{author}{\bibfnamefont{T.}~\bibnamefont{Aschebrock}} \bibnamefont{and}
  \bibinfo{author}{\bibfnamefont{S.}~\bibnamefont{K\"ummel}},
  \bibinfo{journal}{Phys. Rev. Res.} \textbf{\bibinfo{volume}{1}},
  \bibinfo{pages}{033082} (\bibinfo{year}{2019}),
  \urlprefix\url{https://link.aps.org/doi/10.1103/PhysRevResearch.1.033082}.

\bibitem[{\citenamefont{Gontier}(2013)}]{Gontier13}
\bibinfo{author}{\bibfnamefont{D.}~\bibnamefont{Gontier}},
  \bibinfo{journal}{Phys. Rev. Lett.} \textbf{\bibinfo{volume}{111}},
  \bibinfo{pages}{153001} (\bibinfo{year}{2013}).

\end{thebibliography}

\end{document}